\documentclass[a4paper,11pt]{article}
\pdfoutput=1
\usepackage{jheppub}
\usepackage[T1]{fontenc} 
\usepackage{amsmath,amssymb,graphicx,bbm,mathrsfs,nicefrac}
\usepackage{tikz}
\usepackage{amsmath}
\usepackage{amssymb}
\usepackage{graphicx,textcomp,float,gensymb,wrapfig, enumitem,comment,dsfont,framed,slashed,appendix,wrapfig,xcolor}
\usepackage{ bbold }
\usepackage{braket} 
\usepackage{ mathrsfs }
\usepackage[small]{caption}
\usepackage{subcaption}
\usepackage{etoolbox,hyperref,makecell}
\usepackage{feynmp}
\usetikzlibrary{arrows.meta}
\usepackage{empheq}
\usepackage{braket}

\patchcmd{\abstract}{\null\vfil}{}{}{}
\newcommand{\bea}{\begin{eqnarray}}  
\newcommand{\eea}{\end{eqnarray}}

\newcommand{\twolines}[2] {\begin{tabular}{@{}c@{}}#1 \\ #2\end{tabular}}
\usepackage{hhline}
\usepackage{float}
\usepackage{caption}

\usepackage[utf8]{inputenc}

\title{Signatures of Mirror Stars}

\author{David Curtin and}
\author{Jack Setford}

\affiliation{Department of Physics, University of Toronto, Canada}

\emailAdd{dcurtin@physics.utoronto.ca}
\emailAdd{jsetford@physics.utoronto.ca}

\date{\today}

\abstract{
Motivated by theories of Neutral Naturalness, we argue that \emph{Mirror Stars} are a generic possibility in any hidden sector with analogues of Standard Model (SM) electromagnetism and nuclear physics. We show that if there exists a tiny kinetic mixing between the dark photon and the SM photon, Mirror Stars capture SM matter from the interstellar medium, which accumulates in the core of the Mirror Star and radiates in the visible spectrum. This signature is similar to, but in most cases much fainter than, ordinary white dwarfs. We also show for the first time that in the presence of captured SM matter, a fraction of dark photons from the core of the Mirror Star convert directly to SM photons, which leads to an X-ray signal that represents a direct probe of the properties of the Mirror Star core. These two signatures together are a highly distinctive, smoking gun signature of Mirror Stars. We show that Mirror Stars could be discovered in both optical and X-ray searches up to approximately 100-1000 light years away, for a range of well-motivated values of the kinetic mixing parameter.
}

\begin{document}

\maketitle

\newpage

\section{Introduction}
\label{section:intro}

With the conspicuous absence of new physics at colliders to date, the hierarchy problem remains as problematic and urgent as ever. With most canonical models of physics beyond the Standard Model under experimental tension, the key challenge is to find well-motivated models with testable signatures that could have escaped searches to date. 
Dark Matter (DM) searches have also yet to bear fruit; however most constraints on dark matter models generally assume a simplistic dark sector in which all of the dark matter is comprised of a single species. Appealing though this possibility is, it is not clear that it is the most well-motivated scenario. 
Perhaps more compelling is the idea that the dark sector could be non-minimal, in the sense of consisting of more than one stable relic with nontrivial interactions, or a spectrum of composite states \cite{Khlopov:2008ty, Alves:2009nf, Kribs:2009fy, Khlopov:2011tn, Carmona:2015haa}. If the dark sector is related to the Standard Model (SM) by any kind of symmetry, which would be the case if the new physics is connected to the hierarchy problem as in the Minimal Twin Higgs~\cite{Chacko:2005pe},
 then the complexity of the Standard Model itself might motivate a non-minimal dark sector. Understanding the possible signatures of \emph{Dark Complexity} is therefore an important and timely challenge.

There has been significant progress along these lines in recent years; for instance models of DM featuring additional interactions or a small number of states~\cite{Zurek:2013wia,Petraki:2013wwa, 
Alexander:2016aln,Battaglieri:2017aum,
Krnjaic:2017tio,Tulin:2017ara,Gluscevic:2019yal,DeRocco:2019njg,Gresham:2018rqo,Chang:2018rso,Fan:2013yva,Fan:2013tia,McCullough:2013jma,Agrawal:2016quu,Agrawal:2017rvu,Agrawal:2017pnb,
Acevedo:2019gre,Bhoonah:2018gjb,Bramante:2016rdh}. Although these approaches capture some of the  possibilities of Dark Complexity, their relative simplicity still limits the range of phenomena that can be explored.  More complicated dark sector models~\cite{Chacko:2018vss, Berlin:2018tvf,Grossman:2017qzw,Kuflik:2015isi,Hochberg:2014dra,Brax:2019koq,Kribs:2018ilo,Cheng:2018vaj,Renner:2018fhh,Garcia:2015loa, Craig:2015xla, Garcia:2015toa, Farina:2015uea,Hochberg:2018vdo,
Zurek:2013wia,Petraki:2013wwa,
Khlopov:1989fj,Foot:1999hm,Foot:2000vy,Berezhiani:2003xm,Foot:2004pa,Berezhiani:2005ek,Foot:2014mia,Foot:2014uba,Foot:2014osa,Foot:2016wvj,Michaely:2019xpz,
Detmold:2014qqa,Detmold:2014kba,Krnjaic:2014xza,Agrawal:2014aoa,Blanke:2017tnb,Blanke:2017fum,Terning:2019hgj} have been examined, but in general the study of Dark Complexity is made daunting by the vast multitude of possibilities, making it difficult to make concrete physical predictions or identify the most motivated scenarios. 

One way of making progress in the study of Dark Complexity is to consider those dark sectors that are related to the SM by symmetry. Hence the idea of a \emph{mirror} sector~\cite{Chacko:2005pe, Holdom:1985ag, Kolb:1985bf, Glashow:1985ud, Foot:1995pa}. A mirror sector is directly related to the SM via a discrete symmetry like a $Z_2$ symmetry, which, depending on the model, may be more or less exact. The symmetry would dictate that states in the mirror sector analogous to SM matter are charged under a copy of the Standard Model gauge group: $SU(3)_c' \times SU(2)_L' 
\times U(1)_Y'$, though the parameters and masses involved may be different. 
The relation of these mirror sectors to the SM makes it possible to compute a large range of complex physical predictions. Despite these similarities, the physical realizations of  dark sector dynamics can be greatly modified compared to the visible sector if the $Z_2$ symmetry is not exact. These close cousins of the SM sector have not received much attention as dark sector candidates so far.

As it turns out, dark mirror sectors are not only quite predictive but also fundamentally motivated, since they play a central role in theories of Neutral Naturalness, in particular the Mirror Twin Higgs (MTH)~\cite{Chacko:2005pe,Barbieri:2005ri,Chacko:2005vw}.
These models solve the little hierarchy problem by protecting the Higgs via a discrete symmetry that gives rise to one (or more~\cite{Craig:2014aea}) dark mirror sectors. 
The new top partner states that regulate the top contributions to the Higgs mass at the TeV scale are therefore charged under a mirror-QCD force, rather than SM QCD. 
This means that Neutral Naturalness, unlike e.g. minimal supersymmetry, does not generate colored top partner signatures, making these theories compatible with LHC exclusions~\cite{Aaboud:2017ayj, Aaboud:2017nfd, Aaboud:2017ejf, 
Aaboud:2016tnv, Sirunyan:2017pjw, Sirunyan:2017cwe, 
Khachatryan:2017rhw}. Notably, this does not simply remove all signatures of naturalness -- novel collider or cosmological signatures are of the essence.

The original minimal MTH model~\cite{Chacko:2005pe}  is perhaps the most appealing of these theories. The mirror sector is related to the SM by a $Z_2$ symmetry that is only softly broken by a higher scale of electroweak symmetry breaking in the mirror sector, giving $v_B/v_A \sim 3-5$ at the cost of a modest tuning $\sim (v_B/v_A)^2$, and predicting new exotic decays of  the 125 GeV Higgs boson into the mirror sector with $\mathrm{Br} \sim (v_A/v_B)^2$. Such decays are not excluded by LHC data at the $\mathcal{O}(10\%)$ level, motivating $v_B/v_A \gtrsim 3$~\cite{Burdman:2014zta}.
At a scale above several TeV, the model can be UV-completed in supersymmetric, extra-dimensional or composite frameworks~\cite{Falkowski:2006qq,Chang:2006ra,Craig:2013fga,Katz:2016wtw,Badziak:2017syq,Badziak:2017kjk,Badziak:2017wxn, Geller:2014kta,Barbieri:2015lqa,Low:2015nqa, Asadi:2018abu}.
Unfortunately, the minimal MTH theory gives rise to an unacceptable cosmological history, since the mirror and visible sectors are kept in thermal equilibrium by SM-mirror Higgs mixing until temperatures of a few GeV~\cite{Barbieri:2005ri}, resulting in a very large $\Delta N_{eff}=5.7$ 
\cite{Chacko:2016hvu,Craig:2016lyx}, in conflict with current bounds~\cite{Aghanim:2018eyx}
Fortunately, there are two simple solutions to this inconsistency, which are illustrative of the collider-cosmology complementarity of hidden sector experimental signatures.

One approach is hard breaking of the $Z_2$ symmetry to various degrees to remove the light degrees of freedom~\cite{Farina:2015uea,Barbieri:2016zxn,Csaki:2017spo,Barbieri:2017opf}, the most extreme case of this being the Fraternal Twin Higgs~\cite{Craig:2015pha}, which preservers only the minimal third generation mirror components necessary to stabilize the Higgs. These families of theories make the mirror sector unstable, thereby eliminating  cosmological problems, but in doing so give rise to spectacular Long-Lived Particle (LLP) signatures that can be effectively probed at the LHC~\cite{Curtin:2015fna,Csaki:2015fba, Curtin:2018mvb}. Various thermal or asymmetric DM scenarios can arise or be embedded in these theories~\cite{Craig:2015xla,Garcia:2015loa,Garcia:2015toa,Prilepina:2016rlq,Freytsis:2016dgf}. 
 
 Another possibility is that the cosmological problems are solved by dilution of the mirror sector abundance. This is the idea of the \emph{asymmetrically reheated Mirror Twin Higgs framework}~\cite{Chacko:2016hvu,Craig:2016lyx}. Following a period of matter domination after decoupling of the two sectors, a late-time decay that favors the visible sector instead of the mirror sector can naturally lower the temperature of the mirror sector relative to the SM, reducing $\Delta N_{eff}$ below current bounds but not out of future observational reach. A particularly predictive  variant is the $\nu$MTH~
 \cite{Chacko:2016hvu}, in which the decay of GeV-scale right-handed neutrinos dilutes the mirror sector by $v_A^2/v_B^2$, allowing a nonzero detection of $\Delta N_{eff}$ to be be correlated with a nonzero detection of $\mathrm{Br}(h\to \mathrm{invisible})$ due to exotic Higgs decays into the invisible stable mirror sector.\footnote{For a related approach, which achieves dilution via kinetic mixing with a massive mirror photon, see \cite{Harigaya:2019shz}.}

 The asymmetrically reheated Mirror Twin Higgs framework provides an excellent case study of Dark Complexity that is predictive ultimately  because of its connection to a fundamental puzzle, in this case the hierarchy problem.  
If baryogenesis occurs in both sectors, a subdominant fraction of DM would be made up of mirror baryons, giving rise to rich cosmological signals in the Cosmic Microwave Background and Large Scale Structure~\cite{Chacko:2018vss}. 
The mirror baryons in our galaxy could also be observed in DM direct detection experiments, which would provide information on the particle content of the mirror sector as well as the distributions of mirror baryons in our galaxy \cite{MTHastro}.
Finally, these mirror baryons could cool and clump to form  \emph{Mirror Stars}, fusing mirror nuclei and shining in mirror photons in analogy to our SM stars.

The MTH example illustrates that 
Mirror Stars are a striking and generic consequence of Dark Complexity, which can arise whenever the dark sector features analogues of electromagnetism and nuclear physics.
The possibility that some fraction of DM could form Mirror Stars is extremely intriguing.
Their precise distribution in our galaxy is difficult to predict in detail, since mirror-baryonic-feedback processes during galaxy formation make the collapse of the mirror halo even more formidably complicated than the SM visible halo~\cite{MTHastro} (see \cite{Chang:2018bgx} for a study of this in a simplified model with only a dark electron and photon). However the observational signatures of Mirror Stars, once understood, could be a generic probe of a wide class of Dark Complexity scenarios.

Mirror Stars have been discussed before in the context of an \emph{exact} mirror sector (i.e. $v_A = v_B$)~\cite{Foot:1999hm, Foot:2000vy, Berezhiani:2003xm, Mohapatra:1996yy, Mohapatra:1999ih, Berezhiani:2005vv}, but there has not been a careful discussion of their general nature or a concrete estimate of their visible signatures. 
It is therefore our aim to determine whether Mirror Stars have a signal in SM photons that could be detected in astrophysical observations.

We demonstrate in this paper that Mirror Stars lead to spectacular astrophysical signatures if the SM ($A^\mu$) and mirror ($A^\mu_D$) photons have a kinetic mixing $\frac{\epsilon}{2} F_{\mu\nu} F_D^{\mu\nu}$~\cite{Holdom:1985ag}. 
This kinetic mixing does not violate any current cosmological or astrophysical constraints for $\epsilon \lesssim 10^{-9}$~\cite{Chacko:2018vss, Vogel:2013raa}.
In fact, a small mixing parameter is expected to exist, since this renormalizable portal could be generated at any scale and is not forbidden by symmetries. 
For example, the minimal Mirror Twin Higgs, where the low-energy degrees of freedom do not generate this  mixing at up to 3 loop order~\cite{Chacko:2005pe}, may naturally give rise to a 4-loop contribution in the range $\epsilon \sim 10^{-13} - 10^{-10}$~\cite{Koren:2019iuv}.
We show that that such tiny values are exactly of the right order to make Mirror Stars observable.

The detection principle is as follows. As long as  this kinetic mixing exists, Mirror Stars will capture SM matter from the interstellar medium, which quickly accumulates to form a ``SM nugget'' in the core of the Mirror Star. This nugget will be heated up to temperatures $T \sim 10^4$~K by the $\epsilon^2$-suppressed interactions with the mirror stellar matter, which gives rise to an optical signal similar to, but usually much fainter than, standard white dwarfs. Additionally, we show for the first time that thermal mirror photons from the Mirror Star core will undergo \emph{mirror Thomson conversion}, where the captured SM matter acts as a catalyst to convert mirror X-ray photons to  SM X-ray photons. A fraction of these X-rays are able to escape the Mirror Star core, providing a direct window into the Mirror Star interior at an energy scale set by the Mirror Star core temperature. Thus Mirror Stars exhibit a distinctive double signature in both visible and X-ray frequencies, which, if the Mirror Stars are in our stellar neighborhood, can be discovered in both optical and X-ray searches.

In a companion letter~\cite{smallpaper} we show how to estimate these signatures using simplified calculations in the regime in which the SM nugget is optically thin to thermal photons, which covers most of the benchmarks we consider.  In this paper we will present the full calculations and discuss important situations in which the assumptions of \cite{smallpaper} break down, although we emphasize that the simpler calculations are often sufficient for order of magnitude estimates.

We present an overview of the Mirror Star detection principles and an outline of our calculation in Section~\ref{section:overview}.
The signal of Mirror Stars obviously depends not only on the hidden sector particle content and $\epsilon$, but also on the macroscopic properties of Mirror Stars themselves. We will study mirror astrophysics in an upcoming work, but defer the problem for now by defining benchmark Mirror Stars that are based on the SM in Section~\ref{section:benchmark_stars}.
Sections~\ref{section:capture} and~\ref{section:heating_and_cooling} show how to compute the properties of the captured SM matter and its emission spectrum, which are almost completely determined by the Mirror Star mass, age and core temperature. This makes future application of our methods to more general hidden sectors and their Mirror Stars straightforward. 
In Section~\ref{section:results} we apply these methods to our benchmark stars and show they could be discoverable in optical surveys like Gaia and dedicated X-ray observations out to distances in excess of 100 light years. 
We conclude with Section~\ref{section:conclusions}.


\section{Overview}
\label{section:overview}

We start by discussing the physical principles that allow Mirror Stars to be observed, and lay out our calculation to determine their signal. 

\subsection{Mirror Star Detection principle}
\label{section:detection_principle}

The basic principle for the detection of Mirror Stars relies very little on the specifics of the mirror stellar physics. For the purposes of an overview in this section we will consider a dark sector with a dark electron $e_D$, and some form of nuclear physics which converts some dark species $\textrm{H}_D$ into another species $\textrm{He}_D$, providing a source of energy. $\textrm{H}_D$ and $\textrm{He}_D$ will be our mirror `nuclei'. We will further assume that all of these species have charges under the dark $U(1)$ mediated by a massless dark photon.\footnote{Inside the Mirror Star, the dark photon would acquire a small thermal mass~\cite{Knapen:2017xzo}. However, this does not change our discussion below or our physical predictions, since it does not significantly affect the relevant interactions between SM matter and mirror matter.}

Crucial to the detectability of Mirror Stars is the existence of a small mixing between the SM photon and the mirror photon:
\begin{multline}
\label{equation:gauge_mix}
\mathcal L_\mathit{mix} = -\frac{1}{4} F_{\mu\nu} F^{\mu\nu} -\frac{1}{4} F_{D\mu\nu}F_D^{\mu\nu} - \frac{\epsilon}{2} F_{\mu\nu} F_D^{\mu\nu} \\ + \overline e \gamma_\mu(i \partial^\mu + Q A^\mu - m_e)e + \overline e_D \gamma_\mu (i \partial^\mu + Q_D A_D^\mu - m_{e_D})e_D,
\end{multline}
where we show only the electron and dark electron in the interaction Lagrangian. 

As we explain below, there is a basis freedom when dealing with a massless dark photon~\cite{Holdom:1985ag}. Removing the kinetic mixing term in \eqref{equation:gauge_mix} does not uniquely specify a basis for the two photons.  For simplicity we will stick with the basis in which mirror particles with charge $Q_D$ end up with a `millicharge' $\epsilon Q_D$ under the SM photon. In this basis the Lagrangian is, to leading order in $\epsilon$:
\begin{multline}
\label{equation:gauge_unmix}
\mathcal L = -\frac{1}{4} F_{\mu\nu} F^{\mu\nu} -\frac{1}{4} F_{D\mu\nu}F_D^{\mu\nu} \\ + \overline e \gamma_\mu(i \partial^\mu + Q A^\mu - m_e)e + \overline e_D \gamma_\mu(i \partial^\mu + Q_D A_D^\mu - \epsilon Q_D A^\mu - m_{e_D})e_D.
\end{multline}
The bounds on $\epsilon$ for a massless dark photon are model dependent, depending on the mass of the lightest particle charged under the dark $U(1)$. In a MTH scenario in which the mirror electron is 3-5 times heavier than the SM electron, the bounds are approximately $\epsilon \leq 10^{-9}$~\cite{Chacko:2018vss, Vogel:2013raa}. This  comes from both supernovae cooling and requiring that the mirror sector, which is at a lower temperature than the SM sector due to the asymmetric reheating mechanism, is not reheated by interactions with the SM, which would generate large contributions to $\Delta N_{eff}$.

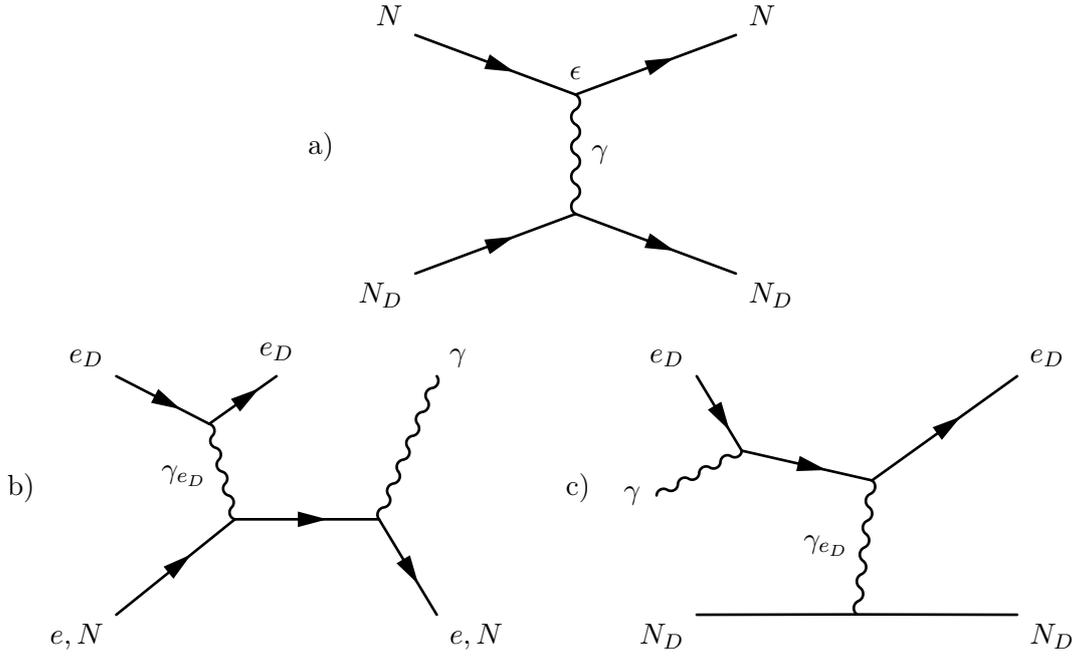
\begin{figure}[t!]
\vspace{5mm}
\centering
a)
\begin{fmffile}{nuclei_scattering}
\begin{tikzpicture}[baseline=(current bounding box.center)]
\node{
\fmfframe(1,1)(1,1){
\begin{fmfgraph*}(150,90)
\fmfleft{i1,i2}
\fmfright{f1,f2}
\fmf{fermion}{i2,b}
\fmf{fermion}{i1,a}
\fmf{photon,label=$\gamma$}{a,b}
\fmf{fermion}{b,f2}
\fmf{fermion}{a,f1}
\fmflabel{$N_D$}{i1}
\fmflabel{$N$}{i2}
\fmflabel{$N_D$}{f1}
\fmflabel{$N$}{f2}
\fmflabel{$\epsilon$}{b}
\end{fmfgraph*}
}
};
\end{tikzpicture}
\end{fmffile} \\
\vspace{10mm}
b) \begin{fmffile}{photon_conversion}
\begin{tikzpicture}[baseline=(current bounding box.center)]
\node{
\fmfframe(1,1)(1,1){
\begin{fmfgraph*}(150,90)
\fmfleft{i1,i2}
\fmfright{f1,f2}
\fmftop{f3}
\fmf{fermion}{i2,a}
\fmf{fermion}{i1,b}
\fmf{photon,label=$\gamma_{e_D}$}{b,a}
\fmf{fermion}{a,f3}
\fmf{fermion}{b,c}
\fmf{fermion,tension=1.5}{c,f1}
\fmf{photon}{c,f2}
\fmflabel{$e,N$}{i1}
\fmflabel{$e_D$}{i2}
\fmflabel{$e,N$}{f1}
\fmflabel{$\gamma$}{f2}
\fmflabel{$e_D$}{f3}
\end{fmfgraph*}
}
};
\end{tikzpicture}
\end{fmffile}
\hspace{5mm}
c) \hspace{2mm} \begin{fmffile}{mirror_free_free}
\begin{tikzpicture}[baseline=(current bounding box.center)]
\node{
\fmfframe(1,1)(1,1){
\begin{fmfgraph*}(150,90)
\fmfleft{i3,i1,i2}
\fmfright{f2,f1}
\fmf{photon}{i1,a}
\fmf{fermion}{i2,a}
\fmf{plain,tension=200}{i3,c}
\fmf{fermion}{a,b}
\fmf{photon,label=$\gamma_{e_D}$}{b,c}
\fmf{fermion}{b,f1}
\fmf{plain,tension=200}{c,f2}
\fmflabel{$\gamma$}{i1}
\fmflabel{$e_D$}{i2}
\fmflabel{$N_D$}{i3}
\fmflabel{$e_D$}{f1}
\fmflabel{$N_D$}{f2}
\end{fmfgraph*}
}
};
\end{tikzpicture}
\end{fmffile}
\vspace{3mm}
\caption{The various important interactions between photons and matter / mirror matter: a) nucleus - mirror nucleus scattering, b) conversion of mirror photons to SM photons via Thomson-like scattering with SM matter, c) absorption of SM photons via free-free absorption, or inverse bremsstrahlung. $\gamma_{e_D}$ is defined to be the state that couples to mirror matter. In the basis of \eqref{equation:gauge_unmix}, $\ket{A^\mu_{e_D}} = \ket{A^\mu_D} - \epsilon\ket{A^\mu}$.} 
\label{figure:diagrams}
\end{figure}

Although the orthonormal basis defined in equation \eqref{equation:gauge_unmix} is the $\ket{A^\mu}$, $\ket{A^\mu_D}$ basis, in which $A^\mu$ couples to $e_D$ but $A^\mu_D$ does not couple to $e$, it is more useful to think in terms of $\ket{A^\mu_e} = \ket{A^\mu}$ and $\ket{A^\mu_{e_D}} = \ket{A^\mu_D} - \epsilon\ket{A^\mu}$. These states do not form an orthonormal basis, but  they represent the states which interact with, and are emitted by, matter and mirror matter respectively. In this picture, interactions between matter and mirror matter are possible because of the overlap between these two states: $|\braket{A^\mu_e | A^\mu_{e_D}}|^2 = \epsilon^2$. As we will see, this picture is helpful to understand which processes do and do not contribute to the Mirror Star signal.

There is an $\epsilon^2$ suppressed interaction between SM nuclei and mirror nuclei, due (in this basis) to exchange of a SM photon (see Figure~\ref{figure:diagrams} a). Crucially, this means that the Mirror Star will be able to \emph{capture} SM matter from the interstellar medium. Although this interaction is small, it can over the course of the star's lifetime lead to substantial accumulation of SM matter in the Mirror Star's gravitational well. Furthermore, once enough material has accumulated it can also \emph{self-capture} more material due to SM-SM interactions which are not suppressed by $\epsilon$.

Via the same interaction shown in Figure~\ref{figure:diagrams} a), the captured SM matter is coupled weakly to the Mirror Star thermal bath; the hot mirror stellar material can heat up the captured material via SM nucleus--mirror nucleus collisions. Since the SM is a `dissipative' sector, there are various, purely SM mechanisms by which it can cool. Since the heating is suppressed while the cooling is unsuppressed, it is not \emph{a priori} clear whether the SM nugget will reach thermal equilibrium with the star. The temperature of the nugget is therefore an important result of our calculation, since it determines the emission spectrum of the nugget and hence the properties of the signal.

Standard Model matter accumulated in the Mirror Star core will also induce \emph{mirror photon conversion}, whereby mirror photons are directly converted into SM photons. This process in shown in Figure~\ref{figure:diagrams} b): a mirror photon emitted by the mirror stellar matter is absorbed by a SM electron and re-emitted as a SM photon. This process is analogous to Thomson scattering, and the cross section is the same up to an $\epsilon^2$ suppression.  
Once converted, the new SM photon $A^\mu_e$ has a low probability of interacting with the millicharged mirror matter, and therefore has a chance of escaping the Mirror Star without being reabsorbed. This means that converted photons can be a direct probe of the Mirror Star \emph{core}. The intensity and shape of this conversion signal is another important property we wish to calculate in detail.

While most SM photons emitted or converted by the captured SM matter escape the Mirror Star in most cases, free-free absorption (or inverse bremsstrahlung) of a SM photon by mirror matter can occur, and is shown in Figure~\ref{figure:diagrams}~c. 
This can attenuate the SM photon signal as it travels through the highly ionized mirror stellar material.\footnote{Importantly, the reverse process $N_D e_D \to N_D e_D \gamma$, with $\gamma$ reaching our telescopes from all parts of the Mirror Star interior, does \emph{not} occur, because a mirror electron will emit a $\gamma_{e_D}$, for which the mirror stellar matter is opaque.}

\begin{figure}[t!]
\centering
\includegraphics[scale=0.6]{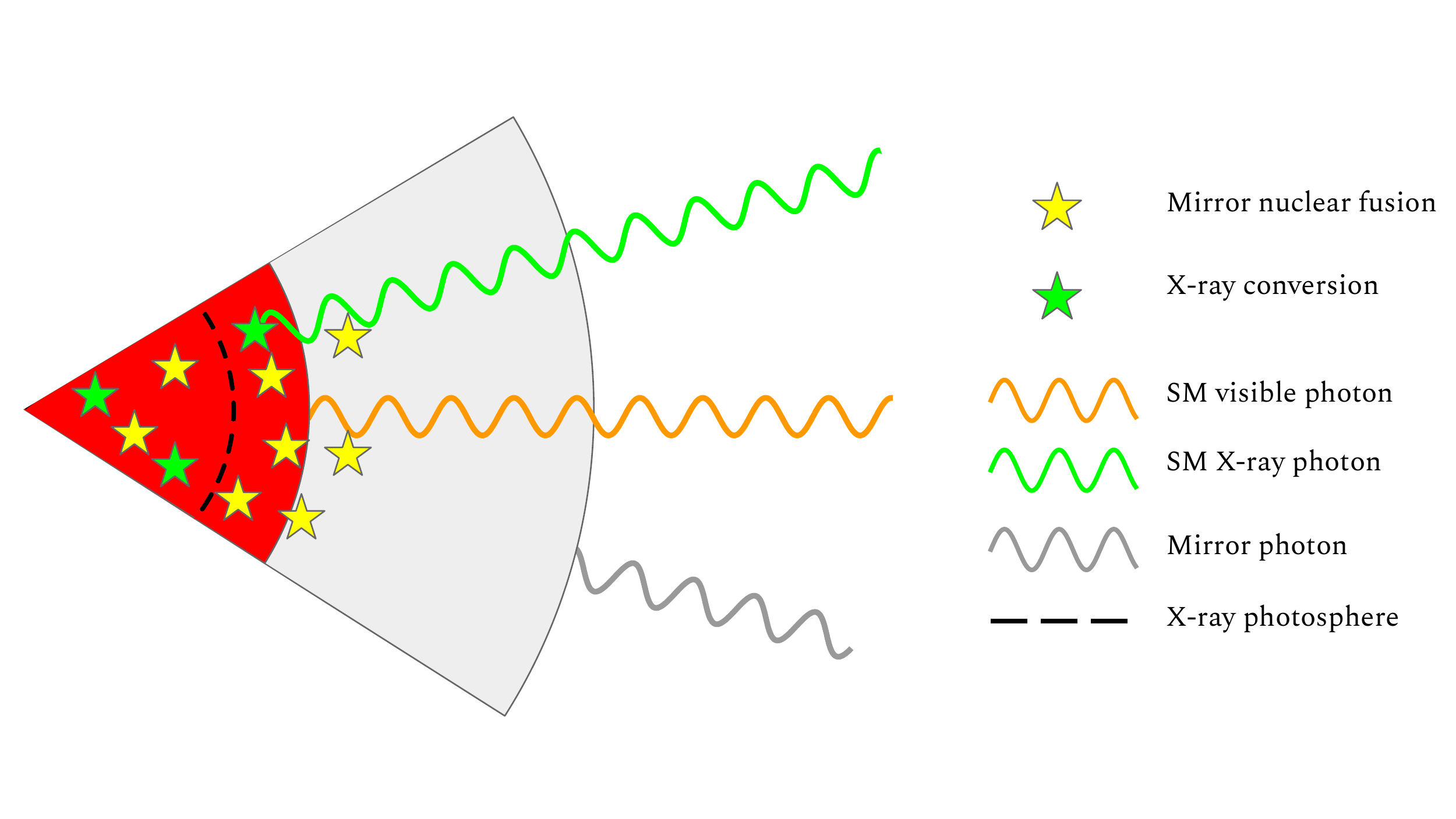}
\caption{Interior of a Mirror Star. Mirror matter is shown grey, captured SM matter in red. The star symbols illustrate why we expect to see X-rays from the Mirror Star core; mirror fusion reactions (yellow stars) produce energy and high energy photons, however these do not escape the star because they only occur in the core. X-ray conversion (green stars) occurs at a constant rate per unit mass of the SM nugget, which is small enough to be contained within the Mirror Star core, so conversions happening near the surface of the nugget will escape the star. Also shown is the $\epsilon^2$ suppressed surface signal, which is expected to be negligible compared to the X-ray and thermal nugget emission.}
\label{figure:mirror_star_wedge}
\end{figure}

We summarise the main observable signals from the Mirror Star in Figure~\ref{figure:mirror_star_wedge}, which are (i) the thermal SM emission of the captured SM material, which as we will show later in this paper is expected to be in the visible part of the spectrum; (2) the emission of \emph{converted} photons from the core of the Mirror Star, which have energies characteristic of the Mirror Star core temperature. If the Mirror Star is sufficiently similar to SM stars, with core temperatures $\mathcal O(10^7\,\textrm{K})$, these photons will in the X-ray spectrum. These X-ray photons will have to escape the SM nugget in order to be observable, and many of them will be absorbed by captured SM atoms before they can escape. 
An important part of our calculation will therefore be determining the fraction of X-rays that escape, and also the shape of the emerging spectrum. For this reason we have indicated in the diagram an X-ray `photosphere', defined such that photons which convert outside this radius will escape the nugget, while those that convert inside are unlikely to escape. 
Crucially, we show that nugget is small enough to be \emph{completely contained within the Mirror Star core} for our benchmark stars. 
This means that the conversion rate is constant per unit mass of SM matter, so there will always be some fraction of X-rays that convert outside the photosphere and are able to escape.\footnote{Compare this to X-ray SM photons in the core of a normal star, which can never `escape' -- fusion rates are not constant per unit mass and take place only at the core of the star, which is well within the photosphere and completely opaque.}

In the same figure we also show the surface emission of the Mirror Star into mirror photons, which is in principle detectable due to the small overlap of the emitted $A_{e_D}^\mu$ state with the SM photon. This means we can potentially observe the surface luminosity of the Mirror Star at $\epsilon^2$ times its mirror luminosity. However, this signal is many orders of magnitude smaller than the signatures of the SM nugget, which are direct SM photon signals.

\subsection{Structure of calculation}

\begin{figure}[t!]
\centering
\includegraphics[width=0.9 \textwidth] {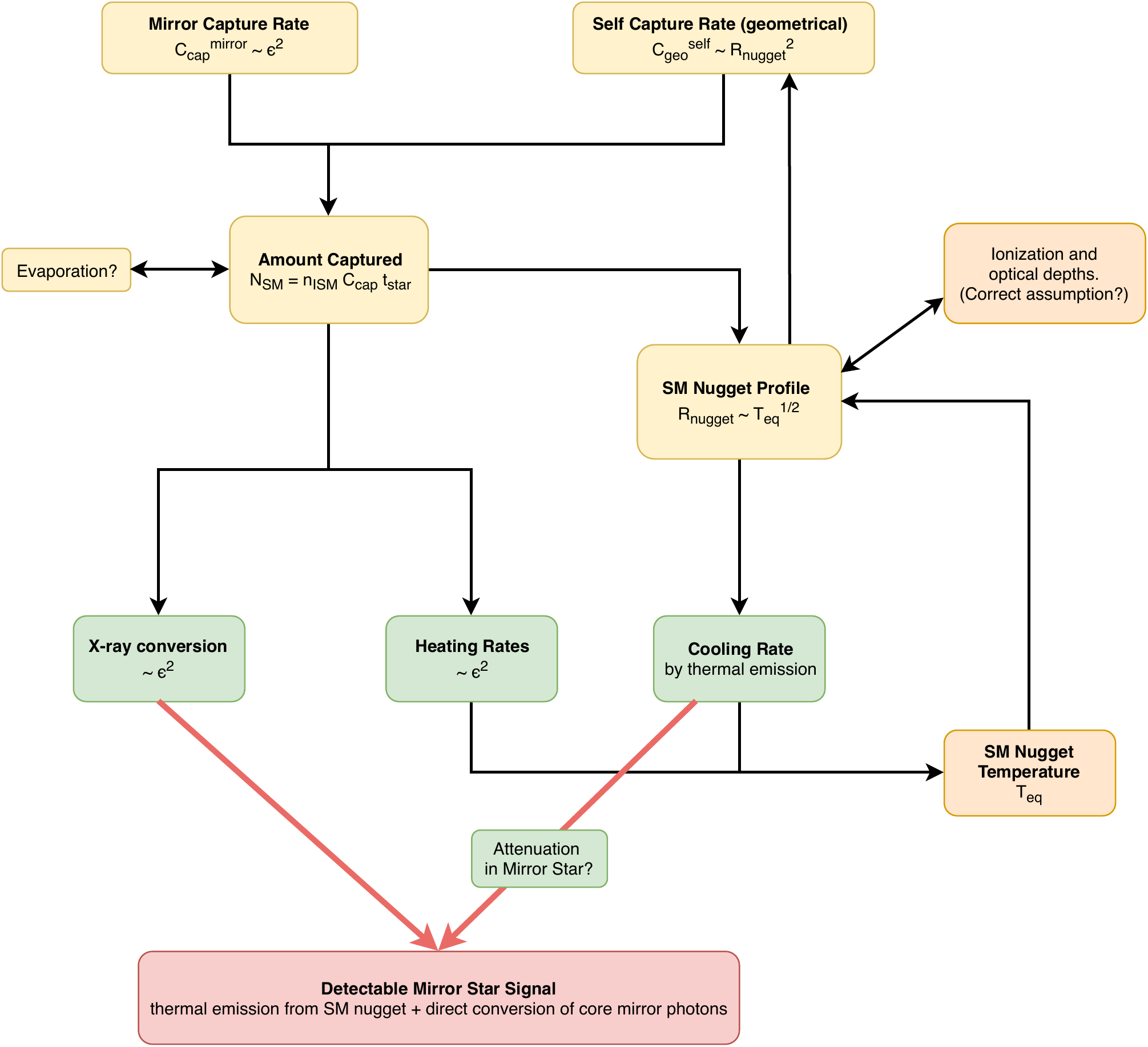}
\caption{
Outline of computation of Mirror Star signal, assuming we know the radial temperature, density and composition profiles of the Mirror Star. 
We also assume that the nugget is either optically thin or thick to thermal photons and check whether this is consistent with the obtained SM nugget profile.
The components in yellow are the focus of Section~\ref{section:capture}, and determine the structure of the nugget (size and density) assuming that we know its equilibrium temperature. The components in green are the focus of Section~\ref{section:heating_and_cooling}, in which we calculate the temperature of the nugget assuming we know its structure. Thus the results of Sections~\ref{section:capture}
 and \ref{section:heating_and_cooling} together allow a consistent solution for all of the nugget's properties. Once the structure is obtained the calculation of the signal is shown in Section~\ref{section:results}.}
\label{figure:flowchart}
\end{figure}

Here we will briefly outline the structure of the calculation and the relevant pieces needed for an estimate of the intensity and spectral shape of the Mirror Star signatures. Figure~\ref{figure:flowchart} shows the different pieces of the computation and their various dependencies, assuming the structure of the Mirror Star (radial profiles of density, temperature and composition) are known. The starting point is the calculation of the capture rate of matter from the interstellar medium. This has two components: mirror capture, which is capture due to scattering of incoming species from mirror nuclei, and self capture, which is capture due to scattering from SM particles that have already been captured. The self capture rate requires knowledge of the properties of the nugget of accumulated matter, in particular the size of the nugget, which depends on its temperature.
On the other hand once we know the temperature of the nugget we can estimate its structure, the level of ionization and whether it is optically thick or thin to photons of different frequencies. This will in turn determine which cooling processes are efficient, and the overal rate of cooling of the nugget. 

Section~\ref{section:capture} is devoted to the calculation of the capture rate and the structure of the nugget, which are both determined assuming we know the \emph{temperature} of the nugget.
In Section~\ref{section:heating_and_cooling} we calculate the heating and cooling rates as a function of the \emph{structure} (size and density), and determine the equilibrium temperature of the nugget. Thus Section~\ref{section:capture} takes as input the results of Section~\ref{section:heating_and_cooling}, and vice versa. We then are able to solve consistently for the total amount captured, the temperature and the structure of the nugget. Once these are known the Mirror Star signal can be calculated, being a function of the properties of the nugget and the known properties of the Mirror Star, such as its core temperature and density. These results are presented in Section~\ref{section:results}.

We emphasize that there are \emph{two} distinctive signatures of a Mirror Star: the thermal emission of the nugget, which is characteristic of the temperature of the nugget; and the X-ray emission from photon conversion, which is characteristic of the temperature of the Mirror Star core. In Section~\ref{section:signal} we will calculate the relative strengths of these two signatures for our benchmark stars and discuss the prospects for their respective detection. We expect that Mirror Stars within observational reach will be close enough for a determination of their parallax, and therefore we will be able to measure the absolute magnitude of the SM photon signal. Thus the smoking gun signature of Mirror Stars would be the faint, but hot, thermal emission of the nugget -- hotter than anything so faint would have any right to be. It may also be possible to resolve the X-ray signature of the converted photons, which would give us a direct probe of the properties of the Mirror Star core and the dark nuclear processes within.

\section{Benchmark Hidden Sector and Mirror Stars}
\label{section:benchmark_stars}

\begin{table}[t]
\begin{center}
\hspace*{-8mm}
\begin{tabular}{| c | c | c | c | c | c | c | c | c |}
\hline
{\Large$\frac{M}{M_\mathit{sun}}$} & \twolines{He \%}{by mass} & {\Large$\frac{R}{R_\mathit{sun}}$} & {\Large$\frac{T_\mathit{core}}{10^7 \,\textrm{K}}$} & {\Large$\frac{n_\mathit{core}}{10^{24} \textrm{cm}^{-3}}$} & {\Large$\frac{L}{L_\mathit{sun}}$} & {\Large$\frac{\tau_\mathit{star}}{\textrm{years}}$} & {\Large$\epsilon_\mathit{crit}^\mathit{mirror}$} & {\Large$\epsilon_\mathit{crit}^\mathit{self}$}  \\
\hline
1 & 0.24 & 1 & 1.54 & 44.9 & 0.96 & $4.3 \times 10^{9}$ & $2.6\times10^{-9}$ & $2.5\times10^{-16}$\\

5 & 0.24 & 3.80 & 2.83 & 6.15 & 721 & $5.6 \times 10^7$ & $5.9\times10^{-9}$ & $3.0\times10^{-15}$\\

50 & 0.24 & 16.1 & 4.13 & 0.74 & $5.18\times10^5$ & $2.3 \times 10^6$ & $1.3\times10^{-8}$ & $8.2\times10^{-14}$ \\
\hline

1 & 0.75 & 1.35 & 2.39 & 20.7 & 13.8 & $1.9\times10^8$ & $4.7\times10^{-9}$ & $4.0\times10^{-16}$ \\

5 & 0.75 & 3.41 & 3.47 & 2.79 & 5570 & $4.1\times10^6$ & $6.7\times10^{-9}$ & $4.5\times10^{-15}$ \\

50 & 0.75 & 20.3 & 4.64 & 0.56 & $1.07\times10^6$ & $5.2\times10^5$ & $2.2\times10^{-8}$ & $1.1\times10^{-13}$ \\
\hline
\end{tabular}
\end{center}
\caption{MESA benchmark stars used in the calculations. Listed are the properties of the stars halfway through their main sequence lifetime. In the second column is the percentage helium composition (by mass) at the beginning of the star's lifetime. The helium mass fraction in the \emph{core} halfway through the star's lifetime is 60.5\% for the low-helium benchamarks, and 84.9\% for the high-helium benchmarks. $\tau_\mathit{star}$ is half the main sequence lifetime, the age of the stars at the time we take them as benchmarks.  In the last columns, $\epsilon_\mathit{crit}^\mathit{mirror}$ ($\epsilon_\mathit{crit}^\mathit{self}$) is the value of $\epsilon$ above which mirror capture is geometric (self-capture becomes geometric within the first 10\% of the star's lifetime), see Section~\ref{section:capture}.
}
\label{table:benchmarks}
\end{table}

The Mirror Star signal will obviously depend on its macroscopic properties like size, age and temperature. Determining the details of this ``mirror stellar astrophysics'' from first principles of the dark sector particle content and interactions is very challenging in generality, and we will pursue this separately in an upcoming work. 
The purpose of this paper is to investigate whether Mirror Stars in general are observable, and how the observational signals follow from the hidden sector and Mirror Star properties. 
We therefore choose to work in a toy model of the dark sector that is an exact copy of the SM. 
That is, we consider a sector with mirror protons, neutrons and electrons with identical masses to the SM proton and electron, with the same charges under the mirror gauge groups, and the same coupling constants. Mirror nuclear fusion allows conversion of mirror hydrogen into mirror helium entirely analogously to the processes taking place in normal stellar cores. This allows us to use standard stellar evolution codes to derive the properties (density and temperature profiles, luminosity, etc.) of our benchmark Mirror Stars.
A SM-like mirror sector is of course a limit of the well-motivated Mirror Twin Higgs model, in which we take $v_B \rightarrow v_A$, although this limit is not favoured phenomenologically. 

We simulate benchmark stars using the stellar evolution code MESA \cite{Paxton2011, Paxton2013, Paxton2015, Paxton2018, Paxton2019}. MESA simulates stars starting from the pre-main-sequence stage, collapsing  an initial spherically symmetric cloud under gravity until nuclear fusion begins and the star enters the main sequence. MESA is able to simulate the entire stellar lifetime, and we define our benchmark stars by their properties halfway through their main sequence lifetime (when 50\% of their core hydrogen fuel has been depleted).
We generate benchmark stars with 24\% helium, corresponding roughly to SM stars. Since in the Mirror Twin Higgs model we expect a significantly higher helium fraction \cite{Chacko:2018vss}, we will also use a set of benchmarks with 75\% helium. We will refer to these benchmarks as the low- and high-helium benchmarks respectively. The mass fraction of hydrogen in the core decreases fairly linearly throughout the star's main sequence lifetime, although the composition of the outer layers is not significantly altered. We verified that the overal capture rates of the benchmarks remain fairly constant throughout the lifetime, so we are justified in treating the capture rates as constant. On the other hand the relative proportions of hydrogen and helium in the core have an $\mathcal O(1)$ effect on the heating rates that we calculate in Section~\ref{section:heating_and_cooling}, and in these calculations we consistently use the benchmarks as defined above, halfway through their lifetime.

The properties of our benchmark stars are summarized in  Table \ref{table:benchmarks}. In our calculations (e.g. of the capture and heating rate) we generally ignore the effects of heavier mirror elements than helium. We also will assume that the presence of the SM nugget in the core of the Mirror Star does not significantly affect the evolution of the Mirror Star. It is clear that if enough SM matter accumulates this assumption must eventually break down, and we will discuss the validity of this assumption with regards to specific benchmarks later in the paper.


\section{Accumulation of SM Baryons in Mirror Stars}
\label{section:capture}

We now outline how to derive the amount of SM matter that accumulates in the core of Mirror Stars due to the kinetic photon mixing interaction, as well as the structure of the resulting SM nugget. We have to consider the capture and evaporation rates, as well as the optical depths of the SM and mirror matter. 

The equilibrium temperature $T_\mathit{eq}$ of the nugget  is a crucial parameter that sets its overall size as well as optical depth at various frequencies. In this section we show how to derive the structure and size of the nugget for a known temperature, and in the next section we show how to derive the equilibrium temperature once the nugget's size and structure is determined. Ultimately, this allows us to iteratively find a consistent solution for the temperature in each of our benchmark scenarios.

\subsection{Capture rates}

The presence of photon mixing automatically implies a non-zero interaction cross section between matter and mirror matter. Thus incoming Standard Model material in the interstellar medium can lose energy via scattering with mirror stellar matter and become caught in the gravitational well of the Mirror Star.

The discussion in this section follows the literature on dark matter capture in the sun or the Earth, since the relevant calculations are analogous \cite{Gould:1987ir, Zentner:2009is, Catena:2015uha, Petraki:2013wwa} especially in the case of asymmetric dark matter \cite{Kaplan:2009ag, Frandsen:2010yj, Iocco:2012wk}, when there is no annihilation channel for the captured species.

The non-relativistic scattering cross section between a nucleus and a mirror nucleus is essentially Rutherford scattering 	 with an $\epsilon^2$ suppression, i.e. the process shown in Figure~\ref{figure:diagrams}~a). The differential cross section can be expressed as
\begin{equation}
\label{equation:cross_section}
\frac{d\sigma}{d E_R} = \frac{2\pi\epsilon^2\alpha^2 Z_1^2 Z_2^2}{m_T v^2 E_R^2}
\end{equation}
where $m_T$ is the mass of the recoiling target, $v$ is the relative initial velocity, and $E_R$ is the target recoil energy.

We will consider capture of both hydrogen and helium, the most abundant species in the Interstellar Medium of our galaxy (ISM). Hydrogen is found in both its ionized and atomic form, while helium is primarily neutral. Since helium is neutral, we will, for completeness, model also the \emph{atom-nucleus} cross section, which at distance scales greater than the Bohr radius of the atom becomes a contact interaction. (This will also be relevant for the heating calculation in Section~\ref{section:heating_and_cooling}.)
Following \cite{Cline:2012is}, we model the interaction between a mirror nucleus and a SM atom as a millicharge scattering off a charge distribution with a screening factor proportional to $\exp(-r/a_0)$, where $a_0$ is the Bohr radius of the outer electron orbital:
\begin{equation}
\label{equation:atom_nucleus}
\frac{d\sigma}{d E_R} = \frac{2\pi\epsilon^2\alpha^2 Z_1^2 Z_2^2}{m_T v^2 (E_R + (2 m_T a_0)^{-2})^2}
\end{equation}
However, as we show below, for velocities typical in the ISM we can ignore atomic form factors in the capture cross section and treat all incoming SM nuclei as fully ionized. (The captured number of electrons is then dictated by charge neutrality, while the ionization of the accumulated SM matter will depend on its equilibrium density and temperature.)

Armed with the scattering cross sections between incoming particles and the target material (mirror or SM), we can define the local capture rate $\Omega_+(w,r)$ as the rate to scatter from a velocity $w$ to a velocity less than the escape velocity (as a function of $r$)~\cite{Gould:1987ir, Zentner:2009is, Catena:2015uha}:

\begin{equation}
\label{equation:capture}
\Omega_+(w,r) = \sum_i n_i(r) w(r) \,\Theta\left( \frac{\chi_i}{\chi_{+,i}^2} - \frac{u^2}{w^2} \right) \int_{E_R^\mathit{min}}^{E_R^\mathit{max}} dE_R\, \frac{d\sigma_i}{dE_R},
\end{equation}
where the sum is over different scattering targets $i$ in the Mirror Star and $n_i(r)$ is the number density of the targets. The integration limits are $E_R^\mathit{min} = \frac{1}{2} m u^2$, the minimum recoil required for the incoming particle to become captured, and $E_R^\mathit{max} = (\chi_i/\chi_{+,i}^2)\frac{1}{2}m w^2$, the maximum possible recoil in a 2-body collision, where $m$ is the mass of the incoming particle, and $w$ is related to the velocity at infinity $u$ via $w(r) = \sqrt{u^2 + v_\mathit{esc}(r)^2}$. Finally we have made use of the definitions $\chi_i = m/m_i$ and $\chi_{+,i} = (\chi_i + 1)/2$. The Heaviside function is zero whenever the $E_R^\mathit{max} < E_R^\mathit{min}$, i.e. when the maximum possible energy transfer is not enough for the incoming particle to be captured, and ensures that the integral does not give a negative contribution to the capture rate in that case.

Given some velocity distribution $f(u)$ and number density $n_{ISM}$  of SM atoms/nuclei in the ISM, the differential capture rate per shell volume is
\begin{equation}
\frac{dC}{dV} = \int_0^\infty du \frac{f(u)}{u} w\,\Omega_+(w,r),
\end{equation}
and the total capture coefficient is
\begin{equation}
\label{equation:volume_integral}
C_\mathit{cap} = 4\pi \int_0^R dr \, r^2 \, \frac{dC}{dV}, 
\end{equation}
with the total capture rate given by 
\begin{equation}
\frac{d N_\mathit{cap}}{d t} = n_{ISM} C_\mathit{cap}.
\end{equation}
For the velocity distribution  we will simply take $u$ to be given by a fixed value, $u = 20$~km/s, which is the approximate velocity dispersion of stars and gas in our local stellar neighbourhood  \cite{Dehnen:1997cq, LopezSantiago:2006xv}. As we now show, the final capture rate scales in a simple way with this velocity, allowing our results to be rescaled to different assumptions about the ISM. 

The escape velocity at the surface of a sun-like star is $v_\mathit{escape} \sim 600$ km/s, and higher for larger stars. We are therefore in the regime where $v_\mathit{escape} \gg u$, and assume that
\begin{equation}
\frac{\textrm{min}(m_i, m_\mathit{SM})}{\textrm{max}(m_i, m_\mathit{SM})} > \frac{1}{4}\left(\frac{u}{v_\mathit{esc}}\right)^2,
\end{equation}
for target nucleus masses $m_i$ and incoming SM masses $m_\mathit{SM}$, which allows us to ignore the Heaviside function. For the case of scattering of hydrogen and helium from mirror hydrogen and helium, this will be satisfied unless the mass hierarchy between the SM and mirror sector is extreme.
For Rutherford-like velocity dependence in the cross section, as in \eqref{equation:cross_section}, the escape velocity then drops out of the capture rate. We find that the capture rate of incoming species $i$ scattering from target species $j$, $C_\mathit{cap}^{i,j}$ simplifies significantly:
\begin{equation}
C_\mathit{cap}^{i,j} \equiv \frac{4\pi N_j \alpha_i^2 Z_i^2 Z_j^2}{m_i m_j u^3},
\end{equation}
where
$m_j$ is the target mass, $m_i$ is the incoming mass, and $N_j$ is the \emph{total} number of scattering targets in the Mirror Star. Note that $\alpha_i$ is equal to $\alpha$ in the case of SM-SM scattering (self-capture), and $\epsilon\, \alpha$ in the case of SM-mirror scattering (mirror capture). We see that the dependence on the escape velocity (as a function of $r$) drops out of the calculation, making the volume integral in equation \eqref{equation:volume_integral} trivial. Strictly speaking we should use $\langle 1/u^3 \rangle$ in the above expression, averaged over the distribution $f(u)$, but we use our benchmark value of $u = 20$ km/s for simplicity, since the final capture rate and hence SM nugget size is easily rescaled.
If we were to take the atom-nucleus scattering cross section in equation \eqref{equation:atom_nucleus}, we would obtain
\begin{equation}
C_\mathit{cap}^{i,j} = \frac{4\pi N_j \alpha_i^2 Z_i^2 Z_j^2}{u \left((1/a_0^j)^2 + m_i m_j u^2\right)},
\end{equation}
where the dependence on the escape velocity again drops out. We also see the condition under which ordinary Rutherford scattering is a good approximation: taking $m_{i,j} \approx m_H$ and $u = 20$~km/s, the effect of the charge screening is in fact small. We therefore treat all captured SM atoms as fully ionized from now on. 

The \emph{overall} capture rate of species $i$ is then
\begin{equation}
\label{equation:overall_rate}
\frac{d N_i}{dt} = n_i^\mathit{ISM} \sum_j \frac{4\pi N_j \alpha_i^2 Z_i^2 Z_j^2}{m_i m_j u^3}
\end{equation}
\begin{equation}
\nonumber
\approx  \frac{1.4\times10^{24}}{\textrm{s}} \left(\frac{n_i^\mathit{ISM}}{\textrm{cm}^{-3}}\right) \sum_j  Z_i^2 Z_j^2  \left(\frac{\epsilon}{10^{-10}}\right)^2
\left(\frac{N_j}{10^{56}}\right)\left(\frac{\textrm{GeV}}{m_i}\right)\left(\frac{\textrm{GeV}}{m_j}\right)\left(\frac{20\,\textrm{km}/\textrm{s}}{u}\right)^3
\end{equation}
 where the sum is over all scattering targets in the Mirror Star, and $n_i^\mathit{ISM}$ is the number density of the incoming species in the interstellar medium. We will only consider capture of hydrogen and helium, which are the two most abundant species in the ISM, and will assume that their average densities over the path of the Mirror Star are given by $n_H^\mathit{ISM} = 1\,\textrm{cm}^{-3}$ and $n_{He}^\mathit{ISM} = 0.1\,\textrm{cm}^{-3}$, which are roughly in accordance with the average values for our galaxy \citep{2011piim.book.....D}.

The sums in equation \eqref{equation:capture} and \eqref{equation:overall_rate} are over all possible targets in the Mirror Star, so include different species of mirror nuclei, and also Standard Model nuclei that have already been captured. It is helpful therefore to split $C_\mathit{cap}$ into a mirror-capture and self-capture component:
\begin{equation}
C_\mathit{cap} = C_\mathit{cap}^\mathit{mirror}(\epsilon) + C_\mathit{cap}^\mathit{self}.
\end{equation}
Since the self-capture rate is proportional to the amount of already captured material, this rate will grow exponentially. This growth, however, will not continue past the so-called geometric limit \cite{Petraki:2013wwa}. There is a geometric limit for both mirror-capture and self-capture; it occurs when all incoming particles traveling through the region in question become captured, so that the capture cross section is essentially given by the physical target area the region presents to the incoming matter. For capture via scattering off mirror matter, the geometric cross section is $\pi R^2$, with $R$ the radius of the star; for self capture, the relevant radius is the approximate size of the region the captured matter accumulates within. This can be estimated using the virial theorem, or by solving for the full structure of the accumulated matter assuming hydrostatic equilibrium (see Section~\ref{section:SMnuggetstructuretheory}). The geometric capture rate is given by \cite{Petraki:2013wwa}
\begin{equation}
\label{equation:geometric_rate}
C_{geo} = \sqrt\frac{3}{2} \frac{\overline v_	\mathit{esc}^2}{u} \pi R^2,
\end{equation}
where $\overline v_\mathit{esc}$ is the average escape velocity. In the case of mirror capture, the average is over the mirror matter distribution in the Mirror Star, while for self-capture, the average is over the SM matter distribution in the nugget. Note that the capture rate in the geometric limit is independent of the inter-particle scattering cross sections, and that the rate is no longer independent of the escape velocity.

Therefore the true capture rate is given by
\begin{equation}
C_\mathit{cap} = \textrm{min}\left( C_\mathit{cap}^\mathit{mirror}(\epsilon), \, C_\mathit{geo}^\mathit{mirror} \right) + \textrm{min}\left( C_\mathit{cap}^\mathit{self}(N_\mathit{cap}), \, C_\mathit{geo}^\mathit{self} \right),
\end{equation}
where $N_\mathit{cap}$ is the amount of SM particles already captured. The geometric limit for self-capture occurs at much lower target densities that for mirror-capture, since the scattering cross section for SM-SM interactions is unsuppressed. 
We define
$\epsilon_\mathit{crit}^\mathit{mirror}$ ($\epsilon_\mathit{crit}^\mathit{self}$) as the value of $\epsilon$ above which mirror capture is geometric (self-capture becomes geometric within the first 10\% of the star's lifetime, assuming only mirror capture in the beginning). Table \ref{table:benchmarks} shows that for the $\epsilon \sim 10^{-12} - 10^{-10}$ range we study, mirror capture (self capture) is always (never) geometric.
Therefore, for the stars we consider, the capture rate of SM species $i$ is in practice given by
\begin{equation}
C_\mathit{cap}^i = \sum_j C^\mathit{mirror}_\mathit{cap, i, j}(\epsilon) + C_\mathit{geo}^\mathit{self} ,
\end{equation}
where $j$ runs over the mirror nuclei, in this case assumed to be just hydrogen and helium.
Since the  self capture rate is time-independent after reaching the geometric limit, the total number SM particles of species $i$ in the nugget is given by
\begin{equation}
\label{equation:Nit}
N_i(t) = n_i^{ISM} C_\mathit{cap}^i t.
\end{equation}
When evaluating the signals of Mirror Stars we set $t$ equal to half the stellar lifetime as a representative example. To a good approximation, the capture rates do not change signficantly throughout the star's main sequence lifetime (even though the core hydrogen/helium fractions change), so we treat the capture rates as constant.

Calculating the geometric capture rate requires knowing size and density of the SM nugget, which depends on its temperature. 
We can preempt the results of Section~\ref{section:heating_and_cooling}. There we demonstrate that for the majority of the benchmark stars we study, the nugget will be optically thin to bremsstrahlung photons, with its cooling rate set by processes that depend on the level of ionization of the nugget. 
We will find that as long as the nugget is optically thin, its equilibrium temperature will be around $4000-7000$~K and it will be isothermal to a very good approximation. The reason for this specific temperature range is that this is the range in which (for the range of densities we encounter), ionization is tiny but sharply increasing in response to higher temperature, so that even when the heating rate varies by orders of magnitude, the cooling rate can compensate by a small change in equilibrium temperature.

Therefore we would be justified in taking a nugget temperature of around $6000$~K  as an ansatz when calculating the capture rate, which gives the correct capture rate to within a factor of two in the optically thin regime. This is the approach we took in the companion letter~\cite{smallpaper}, but we emphasize that in the calculations presented here, we  derive the equilibrium temperature to be consistent with the assumed geometric capture rate.

\subsection{Evaporation rates}

If the mass of the captured particles is sufficiently low, one has to take into account loss due to evaporation. Evaporation is the loss of captured material due to the fact the thermal velocities, either those of the mirror matter or the captured matter, are close enough to the escape velocity of the star that particle collisions can result in ejection of captured matter out of the star.

Evaporation has been studied in the context of dark matter capture in the sun \cite{Griest:1986yu, Petraki:2013wwa}, where one generally finds that above some critical mass, evaporation can be neglected. Because the probability to find a particle above a certain velocity is related to the Maxwell-Boltzmann distribution at temperature $T$, the rate of loss due to evaporation depends exponentially on the mass of the captured species. For instance, equating $\frac{1}{2}m v_\mathit{esc}^2 \sim \frac{3}{2}k T_\mathit{core}$ for the sun, leads to the estimate $m_\mathit{crit} = 
\mathcal O(\textrm{GeV})$. More detailed study suggests a critical mass for the sun of around 3.5 GeV \cite{Griest:1986yu}. For a captured species with mass greater than this, evaporation is never important on stellar lifetimes, while for masses below this, evaporation sets a limit on the amount of material that can accumulate.

This suggests (at least for Mirror Stars of around a solar mass), that evaporation may be important for captured hydrogen, but less so for helium. Evaporation will be less important for more massive stars, since the relative size of the thermal velocities compared to the escape velocity becomes smaller for larger stars, so thermal ejection becomes more difficult.

The local evaporation rate is given by \cite{Griest:1986yu}
\begin{multline}
\Omega^-(r,T,T_N) = \sum_j \, n_j(r) \left( \frac{m}{2\pi k T} \right)^{3/2} \left( \frac{m_j}{2\pi k T_N} \right)^{3/2} \\
\times \int_0^{v_\mathit{esc}} d^3 v \int_0^\infty d^3 v_j \, e^{-m v^2/2 k T}\, e^{-m_j v_j^2/2 k T_N} \, \sigma^i (v, v_j) | v-v_j |,
\end{multline}
where $n_j(r)$ is the number density of different species in the star, $T, T_N$ are the temperatures of the captured matter and mirror nuclei respectively, and
\begin{equation}
\sigma^j(v,v_j) = \int_{E_R^{min}}^{E_R^{max}} dE_R \frac{d\sigma}{d E_R}
\end{equation}
is the cross section for a captured particle to scatter from velocity $v$ up to the escape velocity, as a function of the captured particle's velocity and $v_j$, the velocity of the particle that hits it. The integration limits are $E_R^\mathit{min} = \frac{1}{2}m (v_\mathit{esc}^2 - v^2)$, the minimum amount of recoil to eject the target, and $E_R^\mathit{max}$, which is the same maximum recoil energy as given in equation \eqref{equation:capture}.

The total loss rate due to evaporation is then given by the integral over the stellar volume
\begin{equation}
-\frac{d N_i^{(e)}}{dt} = \int_0^R dr\, n_i(r)\, 4\pi r^2\, \Omega^-(r,T,T(r)),
\end{equation}
where $n_i(r)$ is the density of the captured species. Calculation of the true evaporation rate thus relies on knowing the density and temperature profile of the captured matter. 
We find that evaporation for hydrogen is important if the hydrogen is thermalized with the mirror stellar matter, at temperatures of around $10^7$ K. However, as explained above and derived in the next section, we expect the captured matter to cool to around $10^4$ K. 
For hydrogen temperatures this low, the evaporation rate is in fact negligible.

Matter which is \emph{self-captured}, i.e. captured by direct interaction with the nugget, will very quickly thermalize with the nugget. Evaporation of self-captured matter is therefore negligible, simply because it is never hot enough. Since capture for most of our benchmarks is self-capture-dominated (see Table~\ref{table:nugget_properties}), this simple argument is sufficient to justify neglecting evaporation.

On the other hand, mirror-captured matter will be captured by interactions with the hot mirror stellar material, and there will be a characteristic timescale for it to cool to the temperature of the nugget and sink into the core. Thus the relevant question is to ask how the evaporation timescale for this newly captured, hot material compares to the timescale for it to cool, since it is possible that it could evaporate faster than it can cool down to the temperature of the nugget.

Our approach to answering this question proceeds as follows: assume there is some captured SM matter that has yet to cool to the temperature of the nugget, and that it is uniformly distributed throughout the star with some density $n_\mathit{outer}$. In that case the evaporation loss rate is simply given by 
\begin{equation}
\label{equation:evaprate}
-\frac{d N_i^{(e)}}{dt} = n_\mathit{outer} \int_0^R dr\, 4 \pi r^2\, \Omega^-(r,T(r),T(r)),
\end{equation}
where we have assumed the captured matter is thermalized with the stellar matter. We can then solve for an equilibrium value for $n_\mathit{outer}$ by equating \eqref{equation:evaprate} with the capture rate, assuming that evaporation is the only process that limits the increase in density (for now we are ignoring cooling).

Now, crucially, we can estimate the cooling timescale for this accumulated matter $n_\mathit{outer}$. The timescale for all of the matter to radiate away its thermal energy via bremsstrahlung can be approximated as
\begin{equation}
t_\mathit{cool} \sim \frac{n_\mathit{outer} k T_{star}}{\frac{dP_\mathit{brems}}{dV}(n_\mathit{outer})},
\end{equation}
where the cooling rate $\frac{dP_\mathit{brems}}{dV}$ (which is a function of the density) will be defined in Section~\ref{section:heating_and_cooling}, and we can take $T_\mathit{star}$ to be the stellar core temperature as an overestimate of the cooling timescale\footnote{The timescale to collapse once the material loses pressure support will in general be even faster.}. We can compare this to the evaporation timescale
\begin{equation}
t_\mathit{evap} = \frac{N}{d N^{(e)} / dt},
\end{equation}
where $N$ is the toal number of captured particles. We find that for all the benchmark stars where mirror capture is important compared to self-capture, this cooling timescale is a few orders of magnitude larger than the evaporation timescale, meaning that captured matter cools faster than it can evaporate. We therefore conclude that even for those stars where mirror capture dominates over self-capture, evaporation has a neglible effect on the amount of accumulated matter. This of course should be checked for more general Mirror Stars that may arise in other hidden sector models.

\subsection{Optical depths}

The structure of the nugget and the efficiency of various cooling processes crucially depend on whether the accumulated SM matter is optically thin or thick at various frequencies.
We therefore discuss the mean free path of SM photons in the SM nugget for high frequencies (ionization photons) and low frequencies (bremsstrahlung photons relevant for cooling). 
We find that the nugget is opaque to ionization photons, justifying the use of Saha's equation to solve for the ionization fraction, while the nugget may be transparent or opaque to thermal photons depending on the density.

\subsubsection{Absorption of Ionization Photons}
Cooling processes will only be efficient if the nugget is optically thin to the frequencies of photons emitted via that process. Particularly important is the optical depth to `Rydberg' photons, i.e. photons with energies close to the ionization energies of the constituent species of the nugget. The optical depth to Rydberg photons determines whether the nugget can cool via collisional ionization and recombination. A gas may be optically thin to other frequencies but optically thick to Rydberg photons, since the absorption cross section is strongly peaked near the ionization energy. The photoionization cross section is given by~\cite{1986rpa..book.....R}:
\begin{equation}
\label{equation:photoionizationxsec}
\sigma_\mathit{photo} = \frac{2^5 \pi^2 \alpha^7 m_e^2}{3 \omega^4} \frac{e^{-4(\arctan(\tau))/\tau}}{1-e^{-2\pi/\tau}},
\end{equation}
where $\tau = \sqrt{\omega/\omega_0 - 1}$ and $\omega_0$ is the relevant ionization energy.

For densities, temperatures, and resulting low ionization fractions in our cases of interest, the SM nugget is opaque to photons emitted at ionization energies, as they are efficiently absorbed by neutral atoms. 
This means that collisional  cooling processes are not important, and 
that we can use Saha's equation to self-consistently determine the ionization as a function of temperature and density.

\subsubsection{Absorption and Scattering of Thermal Photons}
\label{section:opticaldepththermal}

We show in Section~\ref{section:heating_and_cooling} that the  equilibrium temperature of the nugget is $\mathcal{O}(10^4 \ \mathrm{K})$. The resulting thermal photons have energies far below ionization threshold. 
Free-free transitions (inverse bremsstrahlung, the equivalent of Figure~\ref{figure:diagrams} c) with all particles in the SM sector) dominate absorption of photons in this energy range. The attenuation coefficient due to free-free absorption is \cite{2011piim.book.....D}
\begin{equation}
\label{equation:free_free}
\kappa_{ff} = \frac{16 \pi^2}{3} \left(\frac{2\pi}{3}\right)^{1/2} \frac{\alpha^3}{m_e^{3/2} T^{1/2} \omega^3} \left( 1 - e^{-\omega/T} \right) Z_i^2\, n_e \,n_i \,g_{ff},
\end{equation}
\begin{equation}
\approx 3.7 \times 10^{-39} \textrm{cm}^{-1} 
\;T_4^{-1/2} \nu_{15}^{-3} \left(1-e^{-4.8\,\nu_{15}/T_{4}}\right) \left(\frac{n_e}{\textrm{cm}^{-3}}\right)\left(\frac{n_i}{\textrm{cm}^{-3}}\right) g_{ff}.
\end{equation}
where $n_e$ is the number density of free electrons and $n_i$ the number density of positive ions, $T_4 = T/10^4 \textrm{K}$ and $\nu_{15} = \nu / 10^{15} \textrm{Hz}$. The attenuation coefficient has units of inverse distance, so that the path length for free-free absorption is given by 
\begin{equation}
\lambda_{abs}^{ff} = \frac{1}{\kappa_{ff}}.
\end{equation}
Free-free absorption peaks strongly at low frequencies and increases with the number density of electrons and positive ions.


Photons in the nugget can also scatter from free electrons, via Thomson scattering, or from neutral atoms, via Rayleigh scattering. At thermal energies, these interactions are almost perfectly elastic, since a photon of energy $E_\gamma$ scattering off a target of mass $m_T$ only loses on average a fraction $E_\gamma/m_T$ of its energy. 
The mean free path for elastic scattering is 
\begin{equation}
\lambda_\mathit{elastic} = \frac{1}{n_e \sigma_\mathit{thoms} + \sum_i n_\mathit{atom}^i \sigma_\mathit{rayleigh}^i},
\end{equation}
where the sum is over different species of neutral atoms in the gas, 
\begin{equation}
\label{equation:thomson}
\sigma_\mathit{thoms} = \frac{8\pi}{3} \frac{\alpha^2}{m_e^2},
\end{equation}
which is the low-energy limit of Compton scattering, valid for photon energies below the electron mass,
and we approximate the Rayleigh scattering cross section by \cite{PhysRev.163.147}
\begin{equation}
\sigma_\mathit{rayleigh}^i = \left(\frac{\omega}{\omega_0^i}\right)^4 \sigma_\mathit{thoms},
\end{equation}
where $\omega_0^i$ is the first ionization energy of atom $i$. 
(The contributions of positive ions to Thomson scattering can be safely neglected due to their much higher mass than electrons.)
If $\lambda_\mathit{elastic}$ is smaller than the size of the nugget $R_{nugget}$, a photon starting from some random point inside the nugget will then undergo a random walk until it either reaches the surface and escapes or is absorbed. The random walk path length to travel the size of the nugget is
\begin{equation}
\label{equation:random_walk}
d \sim \frac{R_{nugget}^2}{\lambda_\mathit{elastic}},
\end{equation}
An approximate condition for the thermal photon to escape the nugget is therefore
\begin{equation}
\label{equation:opticallythincondition}
\lambda_{abs}^{ff} >
\left\{
\begin{array}{lll}
 R_{nugget}  & \mathrm{for} & \lambda_\mathit{elastic} > R_{nugget}
 \\
  \frac{R_{nugget}^2}{\lambda_\mathit{elastic}} & \mathrm{for} & \lambda_\mathit{elastic} < R_{nugget}.
 \end{array}
 \right.
 \end{equation}
 As we now discuss, this condition defines two regimes in which we have to solve for the structure of the SM nugget differently.


\subsection{Solving for the SM Nugget Profile}
\label{section:SMnuggetstructuretheory}

We encounter two regimes when solving for the structure of the accumulated SM nugget. 
For smaller values of $\epsilon$ (around $\epsilon \lesssim 10^{-10}-10^{-11}$ depending on the benchmark), the nugget is transparent to thermal photons and cools via bremsstrahlung emission. 
In this ``optically thin'' regime it is simple to solve for the isothermal SM baryon profile using hydrostatic equilibrium (see below).
At high SM baryon densities, arising for $\epsilon \simeq 10^{-10}$ in some of our benchmark Mirror Stars, the SM nugget is opaque to thermal photons and cools as a black body. In this regime the isothermal assumption breaks down, and we  estimate the SM baryon profile using the virial theorem. 

In either of these cases we can find the nugget profile if the average temperature is known. In practice, we first assume the nugget is optically thin and then make an ansatz for the nugget temperature of $T \sim \mathcal{O}(10^4 \ \mathrm{K})$ to derive a SM nugget profile. For a given profile, the heating and cooling rates can be determined as described in Section~\ref{section:heating_and_cooling} to solve for the equilibrium temperature, and we then iterate towards a consistent temperature solution.
If Eqn.~\ref{equation:opticallythincondition} is not satisfied, we repeat the process under the optically thick assumption.

\subsubsection{Optically thin  regime}
\label{section:opticallythinregimenuggetstructure}

As long as the nugget remains optically thin to bremsstrahlung photons, determination of its size and structure is relatively simple using hydrostatic equilibrium:
\begin{equation}
\frac{d P_{SM}}{d r} = \frac{d}{dr}\left(n_{SM} k T(n_{SM})\right) = - \frac{G M_\mathit{mirror}(r)\overline\mu_{SM} n_{SM}}{r^2},
\end{equation}
where $\overline \mu_{SM}$ is the average mass of SM particles, $M_\mathit{mirror}(r)$ is the Mirror Star mass enclosed as a function of $r$ ignoring the negligible gravitational contribution of the SM baryons, and $P_{SM}$ is the pressure holding up the SM baryons, and we have assumed the ideal gas law holds.\footnote{We assume for simplicity that the SM H and He are well mixed in the nugget. It would be straightforward to solve for the H and He profiles separately, but since H dominates the SM number density and He does not contribute significantly to the ionization fraction at our temperatures of interest, this has only a very small effect on either the SM baryon profile or the computed heating, cooling and emission rates.
}

A crucial result of Section~\ref{section:heating_and_cooling} is that in the optically thin regime, the heating and cooling rates are independent of SM number density. This means that to a very good approximation the nugget has a constant temperature profile, which  simplifies the structure calculation. The solution to \emph{isothermal} hydrostatic equilibrium is given by
\begin{equation}
\label{equation:density_solution}
n_{SM}(r) = C e^{-\int A(r)\, dr}, \;\;\; A(r) = \frac{G M_\mathit{mirror}(r)\overline\mu_{SM}}{k T_\mathit{eq} r^2},
\end{equation}
where $C$ is a constant of integration which is set by the total amount of captured matter, and $T_\mathit{eq}$ is the equilibrium temperature found by equating \eqref{equation:collisional_power} and \eqref{equation:total_emission}.
The local ionization  is found by applying Saha's equation. 
Technically this feeds into Eqn.~(\ref{equation:density_solution}) by affecting the average SM particle mass. Therefore, in full generality, one must choose an input value of $\bar \mu_{SM}$, compute the profile and ionization, and then adjust $\bar \mu_{SM}$ until it is consistent with the found ionization. In practice we reach the optically thick regime before ionization becomes significant, and for the purposes of solving for the SM nugget we can neglect the tiny ionization fraction, fully determining $\bar \mu_{SM}$ in terms of the captured amounts of hydrogen and helium. 

The approximate size of the nugget can be extracted from \eqref{equation:density_solution} by assuming constant density near the Mirror Star core: $M_\mathit{mirror}(r) \approx \frac{4}{3}\pi r^3\rho_{mirror}$, leading to
\begin{equation}
\label{equation:Rvirial}
R_{nugget} \sim \left( \frac{k T_\mathit{eq}}{G \rho_\mathit{mirror} \overline\mu_{SM}} \right)^{1/2},
\end{equation}
which agrees with a naive estimate using the virial theorem, equating thermal kinetic energy with gravitational potential energy \cite{Petraki:2013wwa}. 
In Figure~\ref{figure:density_profile} we plot the density profile for a particular benchmark, with relevant properties summarized in Table~\ref{table:nugget_properties}.

\subsubsection{Optically thick regime}
 
\label{section:opticallythickregimenuggetstructure}

The situation is more complicated if the nugget is optically thick to thermal photons. 
The cooling rate is now no longer a known function of density and temperature, but is set instead by surface emission,  and is a function of the surface temperature and the radius of the nugget:
\begin{equation}
\label{equation:blackbody}
P_\mathit{blackbody} = 4\pi R_{nugget}^2 \sigma_B T_\mathit{surface}^4,
\end{equation}
where 
$T_\mathit{surface}$ is its effective surface temperature (not necessarily equal to its average temperature $T_{eq}$).
In this situation the temperature profile of the nugget will be set by  radiative and/or convective heat transport \cite{1983psen.book.....C}. Solving for this profile, together with the equations of hydrostatic equilibrium, is very analogous to solving the structure equations for a star. 
This requires a much more detailed understanding of the captured material's opacity, which is beyond our scope.

Fortunately, we can obtain a reasonable estimate of the structure and temperature of the optically thick nugget by assuming that it is approximately isothermal, $T_\mathit{eq} = T_\mathit{surface}$ 
with a radius  given by the virial theorem as in Eqn.~\eqref{equation:Rvirial}. We can then solve for $T_\mathit{eq}$ and hence $R_{nugget}$ at equilibrium by requiring the total heating rate to be equal to the rate of black body emission in Eqn.~\eqref{equation:blackbody}.
This essentially assumes that the surface and interior of the nugget are in perfect thermal contact (i.e. by strong convective processes) and therefore gives a \emph{lower} bound on the average temperature $T_{eq}$ and hence a lower bound on the nugget radius, in turn providing an \emph{upper} bound on the nugget surface temperature.

We will be using the above isothermal assumption for all our signal estimates in the optically thick regime. This is justified since, as we now discuss, it underestimates both the thermal and X-ray luminosity of the SM nugget. 

The radius of the nugget scales as $R_{nugget} \sim T_{eq}^{1/2}$. 
Assuming that $T_{eq} \ll T_{core}$, which should be satisfied since cooling processes are not $\epsilon$-suppressed, the heating rate is independent of $T_{eq}$. Therefore, keeping the cooling rate given by Eqn.~\eqref{equation:blackbody} constant gives
\begin{equation}
T_{surface} \sim T_{eq}^{-1/4} \ .
\end{equation}
A higher average temperature would therefore lead to a lower  frequency of the nugget thermal signal, but this dependence is so modest that the nugget would remain observable in optical surveys even for $T_{eq}$ orders of magnitude above the isothermal lower bound. 
On the other hand, the number of accumulated SM particles from geometric self-capture scales as $N_{SM} \sim T_{eq}$. 
That means the  luminosity of the thermal signal would increase with $T_{eq}$, and linearly so if self-capture dominates. 

The X-ray signal, as we discuss schematically in Section~\ref{section:detection_principle}, depends on the fraction of the SM nugget that is close enough to the surface for the converted X-rays to escape. 
The thickness $\Delta R$ of the X-ray photosphere is determined by the competition between the scattering length $\lambda_{scatter}$ of  random-walking X-ray and the characteristic mean free path with respect to absorption via photoionization $\lambda_{abs}$. 
It scales as $\Delta R^2 \sim \lambda_{scatter} \lambda_{abs}$ (see Eqn.~\eqref{equation:opticallythincondition} with $R_{nugget} \to \Delta R$). Both mean free paths scale with the density of the nugget $\lambda \sim 1/\rho$, which scales as $\rho \sim N_{SM}/R_{nugget}^3$. Therefore,
\begin{equation}
\Delta R \sim \frac{T_{eq}^{3/2}}{N_{SM}}
\end{equation}
The number of converted X-rays is proportional to the number of conversion targets $N_{SM}$, whereas the fraction of them that escapes is $\sim \Delta R/R_{nugget}$. The total X-ray luminosity therefore scales as
\begin{equation}
L_{x\textrm{-}ray} \sim N_{SM}  \frac{\Delta R}{R_{nugget}} \sim T_{eq} ,
\end{equation}
confirming that this signal is underestimated by our isothermal assumption as well.

\section{Temperature and Emission Spectrum of Captured SM Baryons}
\label{section:heating_and_cooling}

In the previous section we showed how to find the size and structure of the captured SM nugget in the Mirror Star, assuming a given average equilibrium temperature $T_{eq}$ of the nugget. In this section, we show how to compute that equilibrium temperature for a given SM nugget structure and size, allowing a consistent temperature solution to be found. This naturally leads us to discuss the emission spectrum of the nugget and hence the observational signatures of the Mirror Star.

There are two heating mechanisms, collisions between the SM and mirror nuclei like $p p_D \to p p_D$, and conversion of mirror X-rays in the Mirror Star core into visible X-rays via Thomson conversion $\gamma_D e \to \gamma e$. Collisional heating dominates for our benchmark stars, but it is conceivable that for Mirror Stars with much higher core temperatures than our SM-like benchmarks, X-ray heating plays an important role. 
Equilibrium between collisional heating and cooling via either bremsstrahlung emission (optically thin regime) or black body surface emission (optically thick regime) determines the SM nugget temperature, and therefore the thermal emission spectrum of captured SM baryons, one of the two important Mirror Star signals.
We also discuss how that signal might be attenuated by absorption in mirror matter as it travels out of the Mirror Star, which is an important effect for higher $\epsilon \gtrsim 10^{-10}$.

The second important signal is X-ray conversion. While this process does not appreciably contribute to heating the nugget in our benchmarks, it does allow SM matter to act as a catalyst and convert mirror X-rays directly from the Mirror Star interior into visible X-rays that can escape the star and escape our telescopes. This direct window into the core is an observational smoking gun of Mirror Stars and provides a direct probe of the Mirror Star core temperature, as well as possibly finer details of mirror nuclear physics processes.

\subsection{Heating Rate}

\subsubsection{Collisional heating}

The simplest and in our case most important process that transfers heat from the thermal bath of the mirror matter to the nugget is collisions between mirror nuclei and Standard Model nuclei (see Figure~\ref{figure:diagrams} a).

We can find a simple estimate for collisional heating rate as follows. Assuming that the SM matter is colder than the mirror matter and rejecting its thermal motion, each collision between a mirror nucleus and a SM nucleus will transfer on average $k T_\mathit{mirror}$ worth of energy to the SM nugget. Taking the average relative velocity to be $v_{rel} = \sqrt{3 k T_\mathit{mirror}/\overline m_\mathit{mirror}}$, with $\overline m_\mathit{mirror}$ the average mirror nuclear mass, one can estimate the heating rate per unit volume as
\begin{equation}
\label{equation:naive_heating}
\frac{dP_\mathit{coll}}{dV} = n_\mathit{mirror} n_\mathit{SM} \, v_\mathit{rel}\, \sigma(v_\mathit{rel})\, k T_\mathit{mirror},
\end{equation}
where the cross section is estimated by setting $E_R \rightarrow kT_{mirror}$. However, this significantly underestimates the true heating rate due to the IR enhancement of the scattering cross section. 
A more complete calculation takes a thermal average of the energy transfer as a function of the relative velocities of the colliding particles:
\begin{equation}
\label{equation:collisional_power}
\frac{dP^i_\mathit{coll}}{dV} = n_\mathit{mirror}^i n_\mathit{SM} \left\langle v_\mathit{rel} \int_{0}^{E_R^\mathit{max}} dE_R \, E_R \, \frac{d\sigma}{dE_R}  \right\rangle,
\end{equation}
where  the index $i$ labels different mirror nuclei species, 
$E_R^{max} = 4 \mu^2 v_{rel}^2 /m_{SM}$, $m_{SM}$ stands for $m_\textrm{H}$ or $m_\textrm{He}$, $\mu$ is the reduced mass of the atom and colliding mirror ion, and the cross section is a function of the relative velocity. 
As we show later in this section, we expect the gas to be around $10^4$ K with negligible ionization, so collisional heating is dominated by collisions between mirror ions in the Mirror Star core and SM atoms. 
Substituting in Eqn.~\eqref{equation:atom_nucleus} for the differential cross section leads to
\begin{equation}
\label{equation:real_heating}
\frac{dP^i_\mathit{coll}}{dV} \approx n^i_{mirror} n_{SM} 
\frac{2\pi\epsilon^2\alpha^2 Z_{SM}^2 Z_i^2}{m_{SM} } 
\left\langle
\frac{1}{v_{rel}}
\left(\log\frac{8 \mu^2 v_{rel}^2}{(1/a_0)^2} - 1\right)\right\rangle.
\end{equation}
For $T_{SM} \ll T_\mathit{mirror}$ it is sufficient to assume the relative velocity is given simply by the velocity of the mirror nucleus, and that the recoil energy $E_R$ is the energy gained by the SM nugget. 
The thermal average is straightforward to evaluate, but we note it is well approximated by simply substituting the average relative velocity $v_{rel} = \sqrt{3 k T_\mathit{mirror}/\overline m_\mathit{mirror}}$ into \eqref{equation:real_heating}.
The result is larger than the simple estimate obtained via \eqref{equation:naive_heating} by an order of magnitude due to the log term that appears from the regulation of the integral by the atomic size. 

It is interesting to note that the collisional heating rate \emph{decreases} with increasing mirror star temperature, $P \sim T_{mirror}^{-1/2}$. This is due to the $1/v^4$ dependence in the total scattering cross section $\sigma$, which is very different from the geometric cross section of inter-atomic collisions relevant for thermodynamics at lower temperatures. As the mirror star temperature is increased, heating by X-ray conversion, discussed below, will eventually dominate since it scales with $T_{mirror}^4$.

\subsubsection{Heating by X-ray conversion}

The SM nugget can also draw heat from the photon population in the mirror thermal bath. Mirror photons can scatter off Standard Model charged particles and convert into SM photons, via a Thomson scattering-like process (see Figure \ref{figure:diagrams} b). These high energy photons can then either be reabsorbed by SM matter, scatter elastically, or escape the nugget altogether, depending on the optical depth.

For our SM-like benchmark Mirror Stars, temperatures at the core of the stars are approximately $10^7$ K, so thermal photons are in the X-ray energy range. X-ray energies are significantly higher than the binding energies of hydrogen and helium, so the X-ray photons do not `see' the atomic bound states and will scatter from all charges, whether bound or ionized. Thus the conversion rate does not depend on the amount of ionization of the SM nugget, which will simplify our discussion.

The effective power input available in converted photons is, per unit frequency and volume:
\begin{equation}
\label{equation:conversion_power}
\frac{d^2 P_\mathit{conv}}{dV\,d\nu} = \epsilon^2\, n_{SM} \,\sigma_\mathit{thoms} \,4\pi B_\nu(\nu, T),
\end{equation}
where $B_\nu$ is the Planck spectral radiance function for a black body:
\begin{equation}
B_\nu(\nu, T) = \frac{2 h \nu^3}{c^2} \frac{1}{e^{h\nu/kT}-1}.
\end{equation}
 Note that the Thomson cross section~Eqn.~(\ref{equation:thomson}) is frequency independent. One way of understanding equation \eqref{equation:conversion_power} is to note that $\frac{4\pi}{c} B_\nu(\nu,T)$ is the spectral energy density per unit frequency, i.e. the energy density in photons at frequency $\nu$. One can also think of $n_{SM}\, \epsilon^2 \sigma_\mathit{thoms} \,c$ as an interaction rate; thus the final expression gives the \emph{energy transfer rate} per unit volume.

Performing the integral over frequency one finds the total power per unit volume:
\begin{equation}
\label{equation:total_conversion_power}
\frac{dP_\mathit{conv}}{dV} = \epsilon^2 \, n_\mathit{SM} \, \sigma_\mathit{thoms}\, 4 \sigma_B T_\mathit{mirror}^4,
\end{equation}
where $\sigma_B$ is the Stefan-Boltzmann constant, appearing in the integral over the Planck distribution.

To determine the total amount of heating due to X-ray conversion one needs to calculate what fraction of X-rays are absorbed by the SM material before escaping the nugget, see Section~\ref{section:xrayconversionsignal}. However, we can already compare the two heating rates \eqref{equation:collisional_power} and \eqref{equation:total_conversion_power} to see that even if all of the converted X-rays dump their energy into the nugget, X-ray heating is always expected to be subdominant to collisional heating for our benchmark stars.

The ratio of \eqref{equation:total_conversion_power} to \eqref{equation:collisional_power} is approximately:
\begin{equation}
\frac{dP_\mathit{conv}}{dV}\Big/ \frac{dP^i_\mathit{coll}}{dV} \approx 4\times10^{-4} \frac{1}{Z_{SM}^2 Z_i^2 \beta_\mathit{log}}\left(\frac{10^{24}\textrm{cm}^{-3}}{n_\mathit{mirror}^i} \right)\left( \frac{m_{SM}}{\textrm{GeV}} \right)^{1/2} \left( \frac{T}{10^7 \textrm{K}} \right)^{9/2},
\end{equation}
where $\beta_\mathit{log}$ represents the log correction factor in brackets in equation \eqref{equation:collisional_power}, which takes values roughly in the range 10-14.
This is sufficient to see that the X-ray heating is always subdominant compared to the collisional heating. Values for $n_\mathit{mirror}$ in the core range from around $10^{24}\,\textrm{cm}^{-3}$ to $10^{26}\,\textrm{cm}^{-3}$, and core temperatures are in the range $1\times10^7$~K to $5\times10^7$~K. The above ratio ranges from around $10^{-6}$ for the smallest benchmark star, to around $10^{-2}$ for the largest.
However, if the hidden sector gave rise to Mirror Stars with much higher core temperatures than SM stars, then X-ray heating could dominate.

\subsection{Cooling Rate and Thermal Signal}
\label{section:cooling_rate}

If the nugget is transparent to Rydberg energy photons with a significant population of neutral atoms, it can cool via collisional excitation, collisional ionization and recombination~\cite{1993ppc..book.....P,2011piim.book.....D, Rosenberg:2017qia}. 
However, we find that the nugget is always optically thick to photons at the ionization energy, meaning there are two main possibilities for cooling: if the nugget is transparent to thermal photons with frequency $\nu \sim T_{eq}$ it can cool via bremsstrahlung emission (see Section~\ref{section:opticaldepththermal}), which arises due to a small but non-zero ionization fraction. This is the ``optically thin'' regime discussed in Section~
\ref{section:opticallythinregimenuggetstructure}. If the nugget is opaque to thermal photons, it cools via surface emission of black body radiation. This ``optically thick'' regime was was discussed in Section~\ref{section:opticallythickregimenuggetstructure}. 
For the benchmark stars we consider, both regimes are encountered.

The optically thin regime arises for smaller values of $\epsilon$ or for more massive and hence shorter-lived Mirror Stars. In both cases, the Mirror Stars accumulate less material and the lower-density nugget is transparent to thermal photons. 
The emissivity (power radiated per unit frequency) from bremsstrahlung is \cite{2011piim.book.....D}
\begin{equation}
\label{equation:brems_emission}
j_\mathit{brems}(\nu) = \frac{8}{3} \left( \frac{2\pi}{3} \right)^{1/2} g_{ff,i}\, \frac{\alpha^3}{m_e^2} \left(\frac{m_e}{k T}\right)^{1/2} e^{-h\nu/k T} Z_i^2 n_e n_i,
\end{equation}
where $n_e$, $n_i$ are the number densities of electrons and ions of species $i$, respectively, and $g_{ff,i}$ is the Gaunt factor for free-free transitions, which is classically equal to unity, but in the quantum treatment is a function of frequency and temperature. 
The total integrated power is
\begin{equation}
\label{equation:total_emission}
\frac{dP_\mathit{brems}}{dV} = 4\pi\int_0^\infty j_\mathit{brems}(\nu)\,d\nu = \frac{16}{3} \left( \frac{2\pi}{3} \right)^{1/2} \frac{\alpha^3}{m_e^2} \,(m_e k T)^{1/2} \,\langle g_{ff} \rangle_T\, Z_i^2\, n_e\, n_i,
\end{equation}
where $\langle g_{ff} \rangle_T$ is the frequency averaged Gaunt factor at temperature $T$.

If the density and temperature are such that we are well below the ionization threshold, then, taking pure hydrogen as an example, we find that the solution to Saha's equation for $n_e n_i$ takes the simple form:
\begin{equation}
n_e n_i = n_{SM}\left(\frac{m_e T_{SM}
}{2\pi}\right)^{3/2} \exp\left(-\frac{\omega_0}{T_{SM}}\right),
\end{equation}
where $\omega_0$ is the ionization energy. Thus we see that, as long as we are below the ionization threshold, both the heating (given by \eqref{equation:collisional_power}) and the cooling rates are proportional to $n_{SM}$, so that the solution for the equilibrium temperature becomes independent of density. This means that the nugget is isothermal to a very good approximation, justifying the assumption made in Section~\ref{section:opticallythinregimenuggetstructure}.

In the optically thick regime the nugget cools via surface emission and will approximate a black body. The power output from a spherical black body is given by Eqn.~\eqref{equation:blackbody}.
As discussed in Section~\ref{section:opticallythinregimenuggetstructure}, we assume the optically thick nugget is isothermal with $T_\mathit{eq} = T_\mathit{surface}$, which allows us to solve for the radius by requiring the black body emission power to equal the heating rate of the nugget. This underestimates the thermal and X-ray luminosity, providing a conservative signal estimate.

\subsection{X-ray Conversion Signal}
\label{section:xrayconversionsignal}

Only a small fraction of converted mirror X-rays escape the nugget as SM photons, and they represent only a small fraction of the nugget luminosity. However, these faint X-ray emissions are both detectable and a smoking gun that unambiguously distinguishes the Mirror Star signal from more conventional other astrophysical sources, such as dim white dwarfs. We therefore estimate the X-ray signal carefully.

In analogy to the discussion for thermal photons in Section~\ref{section:opticaldepththermal}, the X-rays can scatter with a mean free path $\lambda_{scatter} \approx (\sum_i n^i \sigma^i_{thoms})^{-1}$ off $i =$ electrons and nuclei, both bound and free. Note that this scattering length is frequency independent. Unlike for thermal photons, we cannot ignore the small energy loss in each collision, which is on average
\begin{equation}
\Delta E^i_\gamma \approx \left( \frac{E_\gamma}{m_i} \right) E_\gamma.
\end{equation}
As discussed in Section~\ref{section:overview}, the \emph{X-ray photosphere} is defined by the depth of the nugget where a converted X-ray will diffuse to the surface before being absorbed. This means the total path length of the random walk has to be less than $\lambda_{abs}(\nu) = (\sum_j n^j \sigma_{photo}^j(\nu) )^{-1}$ for $j = $ neutral atoms, where the photoionization cross section Eqn.~(\ref{equation:photoionizationxsec}) decreases at higher frequencies. This makes the X-ray photosphere frequency dependent. 

To take both energy loss and absorption into account without solving a full diffusion equation for radiation inside the SM nugget, we estimate the X-ray spectrum according to the following expression:
\begin{multline}
\label{equation:xray_spectrum}
\frac{dP_\mathit{x\textrm{-}ray}}{d\nu_\mathit{obs}}=\int_0^{R_\mathit{nugget}}dr \,4\pi r^2 \int_0^\infty d\nu_i\,\frac{\nu_f}{\nu_i}\frac{dP_\mathit{conv}}{dV \,d\nu_i}
\\
\times\Theta\big(\lambda_\mathit{abs}(\nu_f)-N_\mathit{scatter}(r)\lambda_\mathit{scatter}\big) \delta\big( \nu_\mathit{obs} - \nu_f(\nu_i, r) \big).
\end{multline}
This power output per unit frequency per unit volume $d P_\mathit{conv} / dV d\nu_i$ is given by Eqn.~\eqref{equation:conversion_power}. Here $\nu_i$ represents the frequency of a photon when it first converts, and $\nu_f$ is its energy after a random walk of $N_\mathit{scatter}$ of scatters: $\nu_f \approx \nu_i\, m/(\nu_i N_\mathit{scatter} + m)$. The delta function then ensures $\nu_f$ is equal to the observed frequency $\nu_\mathit{obs}$. The Heaviside function accounts for absorption, and ensures that photons do not contribute to the signal if they must travel further than the absorption path length  before escaping the nugget, the absorption path length being a function of frequency.
The number of scatters required to escape is estimated from the density profile as a function of $r$, by assuming the photon must random walk a distance $R_{nugget} - r$, leading to $N_{scatter} = (R_{nugget} - r)^2 / \lambda_{scatter}^2$, where $\lambda_{scatter}$ is the path length for Thomson scattering, assuming that the density along the photon's path remains the same as the density at $r$ (which strictly overestimates both the number of scatters and the chance of absorption, thus \emph{underestimating} the signal). Since the density profile of the nugget is roughly a Gaussian and there is no hard cut off, we take $R_{nugget}$ to be the surface of `last scattering', where the density is such that a photon past this point has a negligible chance of rescattering.

\subsection{Signal Attenuation in Mirror Matter}

In order to observe the signal of Mirror Stars, it is crucial that photons from the SM nugget actually escape the star. 
The escaping SM photons could be absorbed by mirror matter or converted to mirror photons as they pass through the Mirror Star after escaping the nugget. These processes are $\epsilon^2$ suppressed,  but the mirror matter is highly ionized and has high density. We therefore discuss the possible attenuation effect of mirror matter on the Mirror Star signal.

For the X-ray signal, the main concern is Thomson scattering from free charges and re-conversion to mirror X-rays. In a solar mass Mirror Star core, the path length for X-rays to Thomson scatter off \emph{mirror} matter is roughly $\lambda_\mathit{thoms} = 10^{-4} \textrm{m}/ \epsilon^2$ , so that for values of $\epsilon < 10^{-7}$ re-conversion back into mirror X-rays is not a concern.

For the lower energy thermal emission of the nugget we need to also check free-free absorption, which is strongly peaked at low frequencies. We do in fact find that, particularly for the 1 solar mass benchmark with $\epsilon=10^{-10}$, free-free absorption can lead to a significant loss of signal. Even for the other benchmarks, we find significant attenuation in the lower frequency part of the spectrum, although this does not affect the total luminosity output significantly.
The attenuation can be calculated as a function of frequency:
\begin{equation}
\mathcal{I}'(\nu) = \mathcal{I}(\nu) \exp{\left(-\int_0^R \frac{1}{\epsilon^2} \frac{dr}{\lambda_\mathit{abs}(n(r),T(r),\nu)}\right)},
\end{equation}
where $\lambda_\mathit{abs}$ is the path length for free-free absorption (given by the reciprocal of \eqref{equation:free_free}) evaluated with mirror sector parameters as a function of mirror matter density $n(r)$ and temperature $T(r)$. We will use this expression in Section~\ref{section:signal} to find the shape of the thermal emission spectrum and the total emitted luminosity, which is an integral over $\mathcal{I}'(\nu)$.

\section{Results}
\label{section:results}

\begin{table}
\tiny
Hydrogen-rich benchmark stars:
\vspace*{-3mm}
\begin{center}
\begin{tabular}{| c | c | c | c | c | c | c | c | c | c | c |}
\hline
$\frac{M_\mathit{star}}{M_\mathit{sun}}$ & $\log_{10}\epsilon$ & $\frac{M_\mathit{nugget}}{M_\mathit{star}}$ & $\frac{R_\mathit{nugget}}{R_\mathit{star}}$ & $\frac{T_\mathit{nugget}}{\textrm{K}}$ & \twolines{Thick/}{Thin} & \twolines{Self/}{Mirror} & $\frac{L_\mathit{vis}}{L_\mathit{sun}}$ & $\frac{L_\mathit{x\textrm{-}ray}}{L_\mathit{sun}}$ & \twolines{X-ray}{ frac.} & $\chi_{H+}$\\
\hhline{|=|=|=|=|=|=|=|=|=|=|=|}

1 & ${-10}$ & $4.0\times10^{-11}$ & 0.0088 & 75000 & Thick & 0.0098 & 1.7 & $1.2\times10^{-8}$ & 0.014 & -- \\
\hhline{|~|-|-|-|-|-|-|-|-|-|-|}

 & ${-11}$ & $7.9\times10^{-13}$ & 0.0037 & 14000 & Thick & 0.98 & $4.0\times10^{-4}$ & $1.5\times10^{-11}$ & 0.089 & -- \\
 \hhline{|~|-|-|-|-|-|-|-|-|-|-|}

 & ${-12}$ & $2.9\times10^{-13}$ & 0.002 & 5400 & Thin & 40 & $6.7\times10^{-7}$ & $5.0\times10^{-14}$ & 0.16 & $1.0\times10^{-5}$ \\
\hhline{|=|=|=|=|=|=|=|=|=|=|=|}

5 & ${-10}$ & $5.1\times10^{-13}$ & 0.0031 & 18600 & Thick & 0.034 & 0.013 & $1.7\times10^{-7}$ & 0.28 & -- \\
\hhline{|~|-|-|-|-|-|-|-|-|-|-|}

 & ${-11}$ & $1.0\times10^{-14}$ & 0.0017 & 5900 & Thin & 1.1 & $1.9\times10^{-6}$ & $1.2\times10^{-10}$ & 0.92 & $8.0\times10^{-5}$ \\
 \hhline{|~|-|-|-|-|-|-|-|-|-|-|}

 & ${-12}$ & $4.7\times10^{-15}$ & 0.0016 & 5100 & Thin & 94 & $1.1\times10^{-8}$ & $5.3\times10^{-13}$ & 0.94 & $1.1\times10^{-5}$ \\
\hhline{|=|=|=|=|=|=|=|=|=|=|=|}

50 & ${-10}$ & $1.8\times10^{-14}$ & 0.0012 & 6400 & Thin & 0.027 & $2.6\times10^{-4}$ & $8.7\times10^{-7}$ & 0.88 & $6.3\times10^{-5}$ \\
\hhline{|~|-|-|-|-|-|-|-|-|-|-|}

   & ${-11}$ & $5.8\times10^{-16}$ & 0.0011 & 5400 & Thin & 2.3 & $1.4\times10^{-7}$ & $3.1\times10^{-10}$ & 0.97 & $3.1\times10^{-5}$ \\
\hhline{|~|-|-|-|-|-|-|-|-|-|-|}
   
   & ${-12}$ & $3.5\times10^{-16}$ & 0.0010 & 4700 & Thin & 200 & $8.9\times10^{-10}$ & $1.9\times10^{-12}$ & 0.97 & $3.8\times10^{-6}$ \\
\hline
\end{tabular}
\end{center}
Helium-rich benchmark stars:
\vspace*{-3mm}
\begin{center}
\begin{tabular}{| c | c | c | c | c | c | c | c | c | c | c |}
\hline
$\frac{M_\mathit{star}}{M_\mathit{sun}}$ & $\log_{10}\epsilon$ & $\frac{M_\mathit{nugget}}{M_\mathit{star}}$ & $\frac{R_\mathit{nugget}}{R_\mathit{star}}$ & $\frac{T_\mathit{nugget}}{\textrm{K}}$ & \twolines{Thick/}{Thin} & \twolines{Self/}{Mirror} & $\frac{L_\mathit{vis}}{L_\mathit{sun}}$ & $\frac{L_\mathit{x\textrm{-}ray}}{L_\mathit{sun}}$ & \twolines{X-ray}{ frac.} & $\chi_{H+}$\\
\hhline{|=|=|=|=|=|=|=|=|=|=|=|}

1 & ${-10}$ & $1.0\times10^{-12}$ & 0.0050 & 32000 & Thick & 0.025 & 0.033 & $2.9\times10^{-8}$ & 0.23 & -- \\
\hhline{|~|-|-|-|-|-|-|-|-|-|-|}

 & ${-11}$ & $2.1\times10^{-14}$ & 0.0022 & 6300 & Thin & 1.2 & $3.4\times10^{-6}$ & $2.2\times10^{-11}$ & 0.83 & $3.5\times10^{-4}$ \\
\hhline{|~|-|-|-|-|-|-|-|-|-|-|}

 & ${-12}$ & $9.9\times10^{-15}$ & 0.0021 & 5400 & Thin & 98 & $3.3\times10^{-8}$ & $1.1\times10^{-13}$ & 0.90 & $4.5\times10^{-5}$\\
\hhline{|=|=|=|=|=|=|=|=|=|=|=|}

5 & ${-10}$ & $2.2\times10^{-14}$ & 0.0025 & 6900 & Thin & 0.028 & $9.5\times10^{-5}$ & $5.5\times10^{-8}$ & 0.92 & $5.4\times10^{-4}$ \\
\hhline{|~|-|-|-|-|-|-|-|-|-|-|}

 & ${-11}$ & $7.2\times10^{-16}$ & 0.0023 & 5800 & Thin & 2.4  & $1.2\times10^{-7}$ & $1.9\times10^{-11}$ & 0.96 & $2.7\times10^{-4}$ \\
\hhline{|~|-|-|-|-|-|-|-|-|-|-|}

 & ${-12}$ & $4.4\times10^{-16}$ & 0.0021 & 5000 & Thin & 200 & $9.4\times10^{-10}$ & $1.1\times10^{-13}$ & 0.96 & $3.2\times10^{-5}$ \\
\hhline{|=|=|=|=|=|=|=|=|=|=|=|}

50 & ${-10}$ & $2.5\times10^{-15}$ & 0.00091 & 6500 & Thin & 0.047 & $5.2\times10^{-5}$ & $2.1\times10^{-7}$ & 0.96 & $2.1\times10^{-4}$ \\
\hhline{|~|-|-|-|-|-|-|-|-|-|-|}

   & ${-11}$ & $1.2\times10^{-16}$ & 0.00084 & 5500 & Thin & 4.0 & $4.4\times10^{-8}$ & $1.0\times10^{-10}$ & 0.97 & $8.3\times10^{-5}$ \\
\hhline{|~|-|-|-|-|-|-|-|-|-|-|}

   & ${-12}$ & $8.2\times10^{-17}$ & 0.00079 & 4800 & Thin & 340 & $3.2\times10^{-10}$ & $7.0\times10^{-13}$ & 0.97 & $1.0\times10^{-5}$ \\
\hline
\end{tabular}
\end{center}
\caption{SM nugget properties for the benchmarks we consider. Thick/Thin refers to whether the nugget is in the optically thick or thin regime,  Self/Mirror gives the relative importance of self-capture vs mirror-capture ($\frac{dN^{(s)}/dt}{dN^{(m)}/dt}$),  ``X-ray frac.''  is the fraction of total converted X-ray power that escapes the nugget as signal, and $\chi_{H+}$ is the ionization fraction of the hydrogen component of the nugget, which dominates the free electron density. 
For the 1 $M_{sun}$ benchmark and $\epsilon = 10^{-10}$, the nugget luminosity we obtain exceeds that of the star, clearly signaling that our assumption of neglecting the effect of the captured SM matter on the Mirror Star has broken down. }
\label{table:nugget_properties}
\end{table}

In this section we summarize our numerical results, obtained by solving for the SM nugget profile and equilibrium temperature and computing the resulting emission spectrum for all our benchmark stars at various values of $\epsilon$. Most scenarios give rise to an optically thin nugget, in which case the simplified calculation of~\cite{smallpaper} gives comparable results.  

In Table~\ref{table:nugget_properties} we list the properties of the SM nuggets in all of our benchmark cases, including the mass and size of the nugget, whether it is optically thick or thin, and the luminosities of both the optical and X-ray signals.

We find that Mirror Stars generate a detectable and highly distinctive astrophysical signals at optical and X-ray frequencies. We visualize the range of signals generated by our benchmark stars in a Hertzsprung-Russell diagram, see Figs.~\ref{figure:HRplot} and~\ref{figure:HRplot_helium}, and show that such signals could be detected in future or even existing observations.

\subsection{SM Nugget Profiles}
\label{section:structure_of_nugget}

\begin{figure}[ht]
\centering
\includegraphics[scale=0.5]{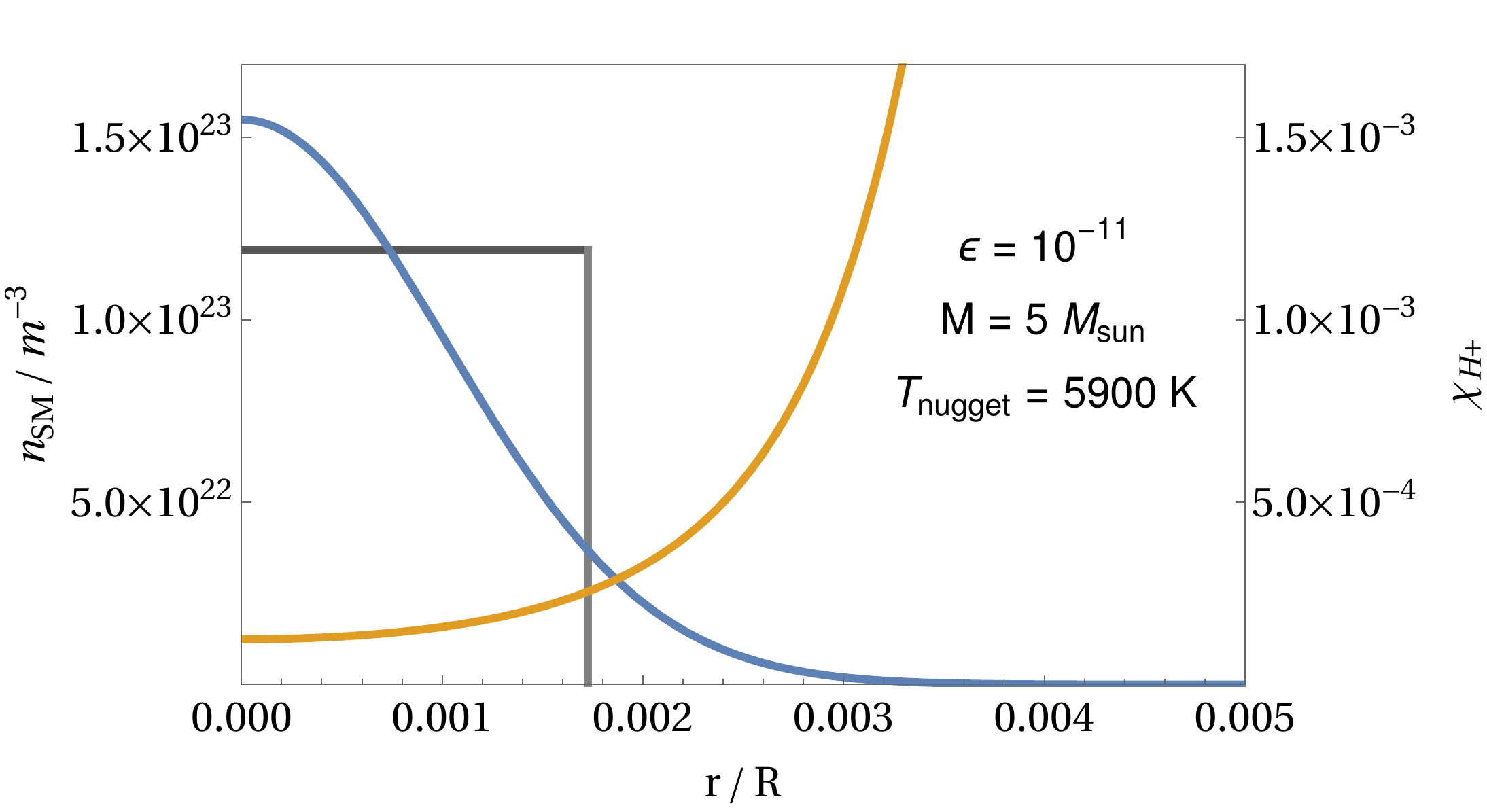}
\caption{SM nugget density profile for a benchmark Mirror Star. In grey we show the profile one would obtain by assuming the nugget has constant density, with a radius given by the virial theorem, as we assume in the more simplified calculation of \cite{smallpaper}. For different values of $\epsilon$, the shape of the profile is essentially the same, its width scaling with $T_\mathit{nugget}^{1/2}$. In orange we show the ionization fraction of hydrogen (which dominates the free electron density); for the majority of the mass of the nugget the variation in the level of ionization is modest, while it increases in the outer regions. The optical depth due to free-free absorption depends only on the total density of free electrons, which strictly decreases from the centre of the nugget outwards and vanishes at larger distances.}
\label{figure:density_profile}
\end{figure}

In the optically thin regime the isothermal profile solution,  given by \eqref{equation:density_solution}, is determined once we know the composition of the nugget and its equilibrium temperature. In Figure~\ref{figure:density_profile} we show the nugget density profile for the 5 solar mass benchmark Mirror Star with $\epsilon = 10^{-11}$. In this case the temperature of the nugget is approximately $ T_\mathit{eq} = 5900$ K and its virial radius is $0.17\%$ the radius of the star, contained deep within the core region. 

The profile we show in Figure~\ref{figure:density_profile} is representative of other optically thin benchmark cases. The profile is very nearly a Gaussian, with a width that scales with $T_\mathit{eq}^{1/2}$, and an overall height which is set by the total amount of material captured.

In the optically thick regime with the conservative isothermal assumption, the nugget radii are generally of order  0.1\% of the star's radius. The temperature of the nugget is a factor of a few higher than in the optically thin case, see Table~\ref{table:nugget_properties}, which is reasonable given it can only cool via surface emission, as opposed to optically thin bremsstrahlung emission from the entire nugget.

We have not obtained explicit solutions for the density profile in the optically thick case, except that the rough size of the distribution is given by the virial radius. 
Fortunately, at our current level of precision, the predictions for the SM nugget emission spectra are quite insensitive to the precise density profile of the nugget:
\begin{itemize}
\item The SM nugget radius, determined self-consistently by its average temperature, does set the geometric self-capture rate, but a rough estimate of $R_{nugget}$ by the virial theorem is sufficient to estimate the number of captured SM nuclei $N_{SM}$ up to a factor of about 2. Since all signals scale linearly with $N_{SM}$, this sets the precision of our luminosity predictions.
\item The thermal emission spectrum (Section~\ref{section:cooling_rate}), which dominates the total luminosity, is completely determined by $N_{SM}$ irrespective of the precise density profile. In the optically thin case, this is because the local SM density drops out when solving for temperature by equating the bremsstrahlung cooling and collisional heating rates. In the optically thick case, it is because the black body emission power is completely determined by size and radius of the nugget with our isothermal assumption. 
\item One might suspect that the precise shape of the density profile is important to determine the escaping X-ray fraction. However, the dependence on profile shape is very modest as here well: substituting the gaussian solution derived in the optically thin regime (blue curve in Figure~\ref{figure:density_profile}) for the simple assumption of a constant SM nugget density (grey curve) has minimal effect on the spectrum shape and changes the overall X-ray luminosity by less than a factor of 2. 
\end{itemize}
In the companion letter~\cite{smallpaper} we focus on the optically thin regime and make the simplifying assumption of constant nugget density to derive the X-ray spectrum. Here, we use the more realistic gaussian profile of Figure~\ref{figure:density_profile}) with width set by the virial radius for the optically thin \emph{as well as the optically thick} cases.
We have not justified this by explicitly solving for the profile shape in the optically thick case, but the signal only depends modestly on the profile shape, and the Gaussian assumption is likely to be a better approximation than constant density.
We leave careful determination of the density distribution in the optically thick case for more detailed future studies.

\subsection{SM Nugget Emission Spectra}
\label{section:signal}

\begin{figure}[t!]
\centering
\includegraphics[scale=0.4]{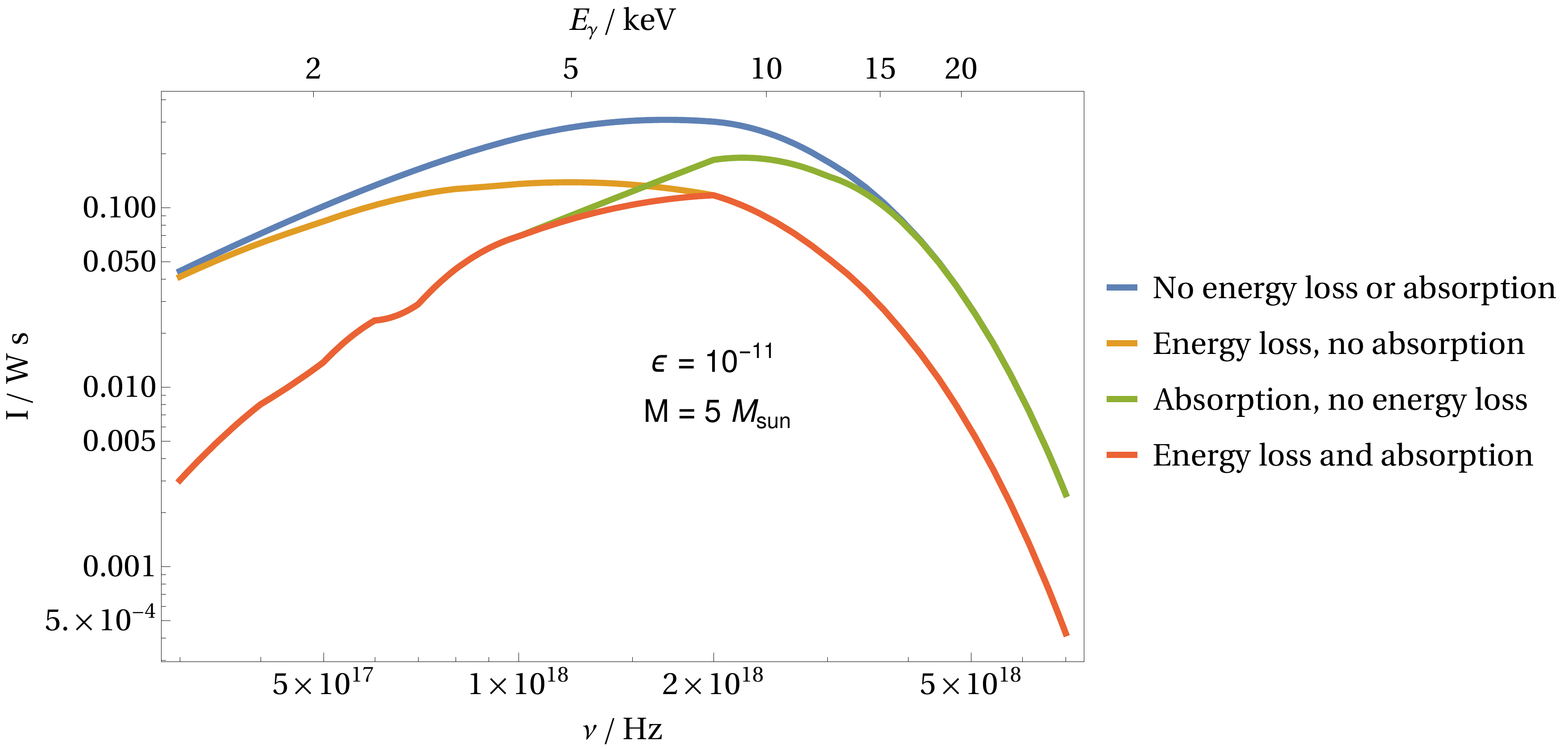}
\caption{Modifications to original X-ray emission spectrum due to energy loss via Thomson scattering and X-ray absorption. Absorption effects due to free-free absorption are stronger at lower frequencies.}
\label{figure:energy_loss_and_absorption}
\end{figure}

\begin{figure}
\centering
\includegraphics[scale=0.38]{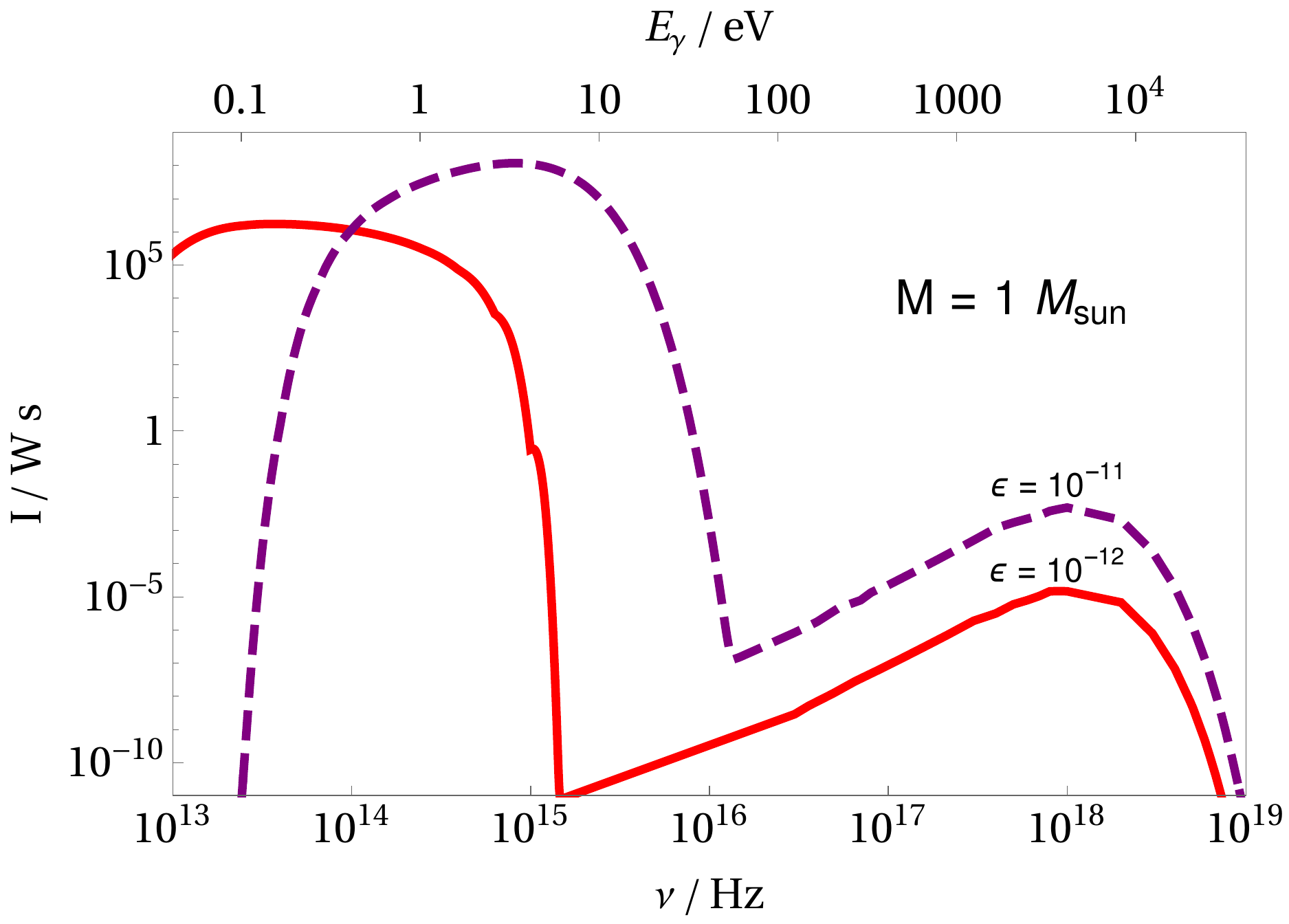}
\includegraphics[scale=0.38]{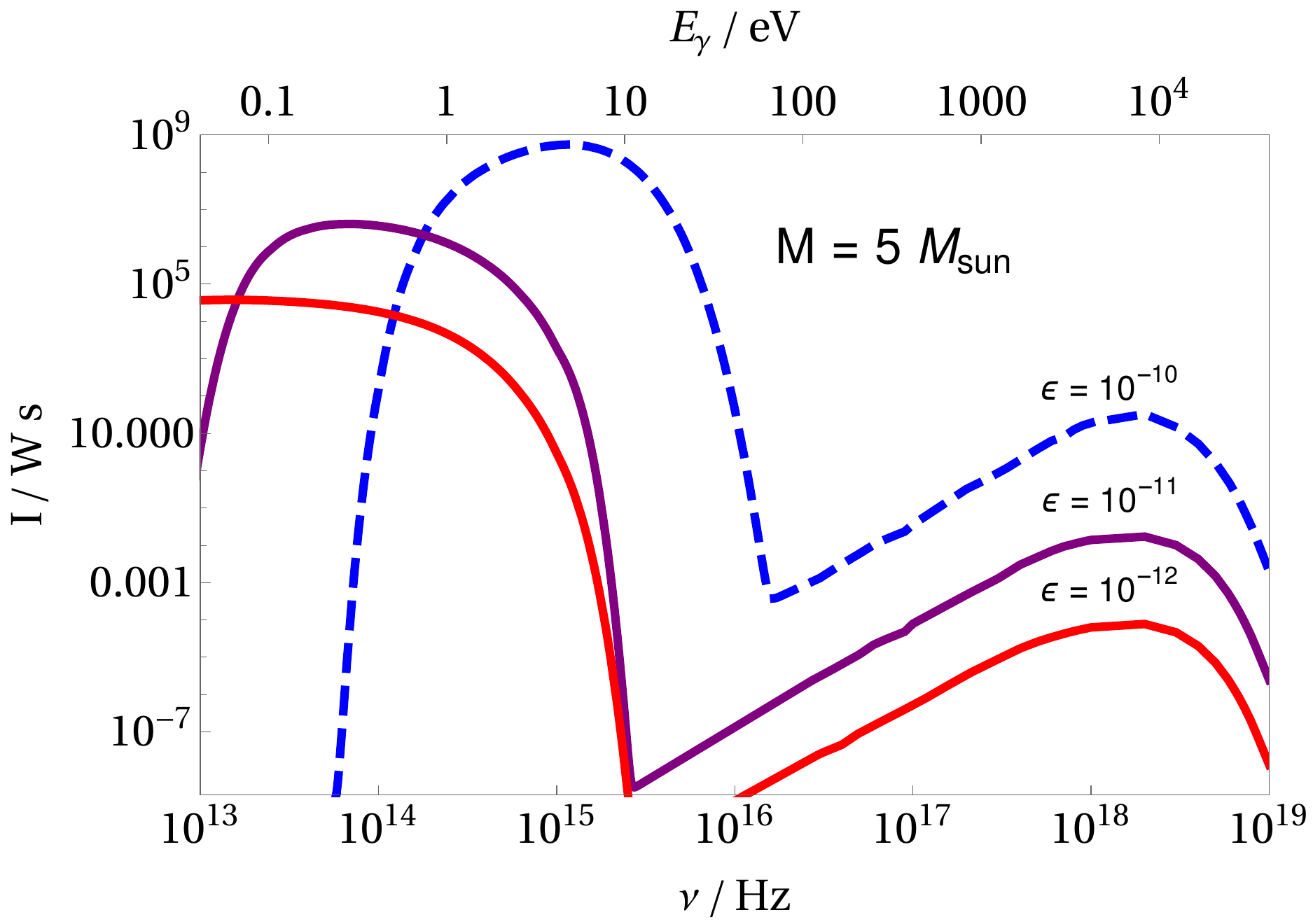}
\\
\includegraphics[scale=0.38]{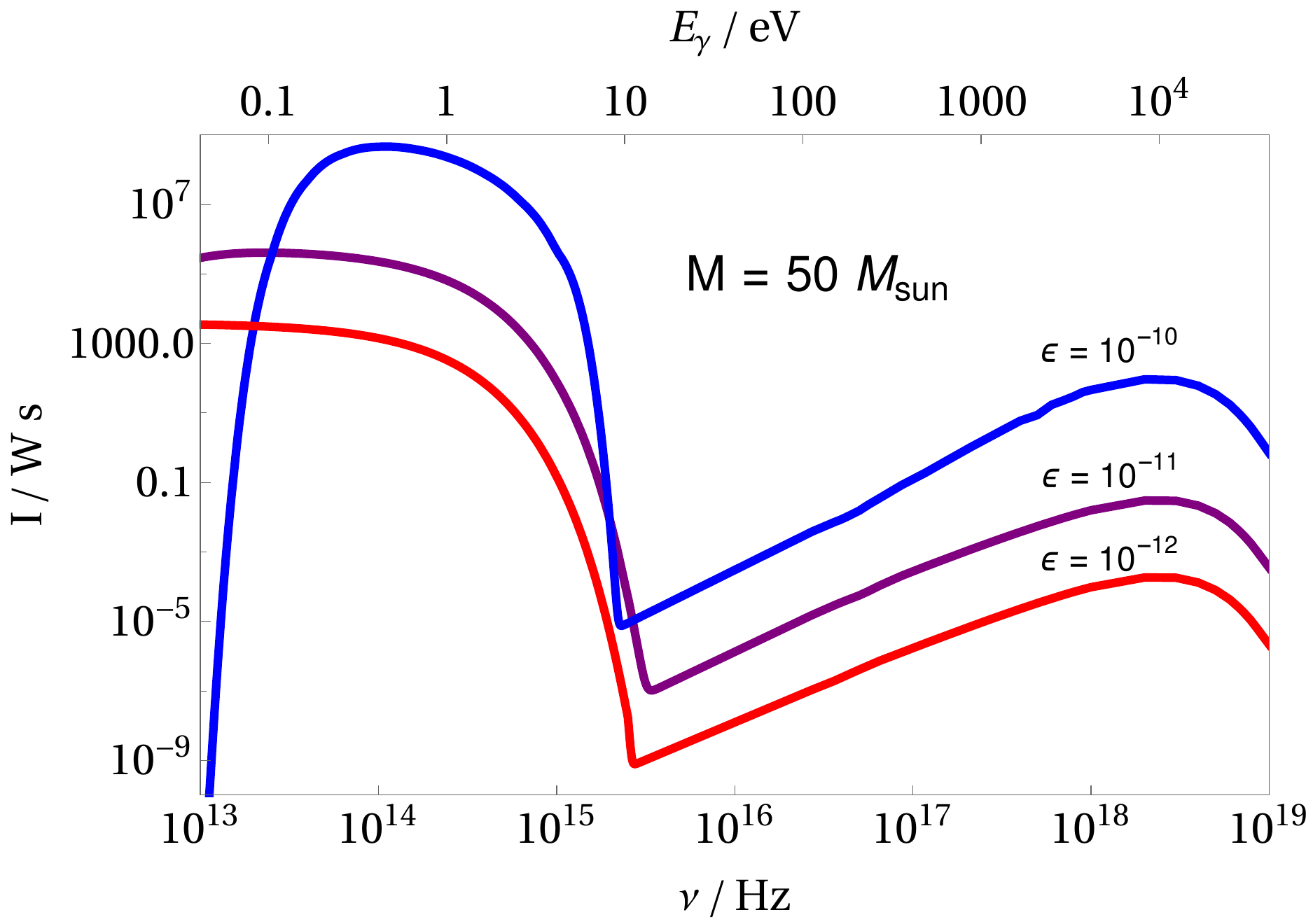}
\captionof{figure}{Nugget emission spectra for all of the benchmarks we consider. We plot the intensity per unit frequency in Watt-seconds as a function of photon energy.  Solid lines indicate benchmarks for which the nugget is optically thin, while dashed lines indicate optically thick benchmarks. Attenuation at low frequencies is due to free-free absorption by mirror electrons, hence is more pronounced at higher values of $\epsilon$. We do not show the spectra from the 1 solar mass $\epsilon=10^{-10}$ benchmark, since, as explained in the text, the total predicted luminosity is greater than the luminosity of the star,
signaling that our assumptions of ignoring the effect of the SM nugget on the Mirror Star itself have broken down}
\label{figure:spectra}
\end{figure}

We now discuss the thermal and X-ray emissions of the SM nuggets for our benchmark scenarios. 

In Figure~\ref{figure:energy_loss_and_absorption} we show the modifications to the original X-ray emission spectrum  due to both energy loss and absorption, for the 5 solar mass  benchmark star with $\epsilon=10^{-11}$. We can see that absorption is significantly more important at lower X-ray frequencies, while for higher frequencies the energy loss in scattering processes is more important. This demonstrates that both effects need to be taken into account in order to obtain the correct final signal shape.

In Figure \ref{figure:spectra} we show results for the full SM photon emission spectra for all of our benchmarks. Characteristic of all of the cases we study is the clear distinction between the two signals, the thermal emission of the nugget and the converted X-ray emission. Since the former has a characteristic temperature of $\sim 10^4$ K (temperature of the nugget) and the latter has a temperature of $\sim 10^7$ K (temperature of the Mirror Star core), the two features in the spectrum are very distinct. 
In the optically thin cases, the thermal emission has the characteristic shape of bremsstrahlung emission, given by \eqref{equation:brems_emission}, which is mostly flat for frequencies below the temperature of the gas. The attenuation of the curve at low frequencies is due to free-free absorption by the mirror stellar matter, which is more important at higher values of $\epsilon$. In the optically thick case, the nugget emission is assumed to be black body, with attenuation in the low frequency part of the spectrum by free-free absorption by mirror matter, and attenuation at X-ray frequencies due to ionization of atoms in the nugget. 
Table~\ref{table:nugget_properties} shows that, as expected, the fraction of converted X-rays escaping the nugget is much higher in the optically thin case than the optically thick case.

It is possible that additional, unaccounted for sources of opacity alter some of our results. Small quantities of heavier elements with smaller ionization energies than hydrogen in the SM nugget may give extra contributions to opacity from photoionization processes. Furthermore, we have neglected the effect of negative hydrogen ions, which have an `ionization' energy of around 0.7 eV, so thermal photons at temperatures $\sim 4000 - 7000$ K have enough energy to liberate the extra electron and be absorbed. Negative hydrogen ions are in fact an important contribution to the opacity of stellar atmospheres~\cite{2014tsa..book.....H}.
Even so, we do not expect this to significantly alter our conclusions. It may be that some of our optically thin benchmarks become optically thick after including these extra sources of opacity. This would not change the total luminosity \emph{per mass} of the nugget, though it may shift the surface temperature. A greater opacity translates to a larger, less dense nugget, which means that more SM matter gets captured and more of the X-ray signal will escape. Therefore, in neglecting these sources of opacity, our present calculation is conservative in that it \emph{underestimates} both the X-ray and the thermal signal.
We intend to revisit these subtleties in a future work.

\subsection{Mirror Star Detection Prospects}

The calculations outlined above allow us to estimate the shape and overall luminosity of the two distinct Mirror Star signatures. A common feature of all our benchmarks is a thermal signature in the visible or near-visible spectrum -- with colour temperatures not very different from those of ordinary stars -- and an X-ray signal characteristic of the core temperature of the Mirror Star, with photon energies in the range 1-10 keV. Compared to standard astrophysical objects, such a signature is spectacularly alien. 
In most cases the optical signal is much too faint to be compatible with its high temperature, due to the nugget 
sitting in the gravitational well of the much more massive, invisible object and being weakly coupled to its (much hotter) thermal bath. 
The dual visible and X-ray signatures suggest an object that somehow has two distinct and widely different temperatures. This signature  is a smoking gun of a Mirror Star. 

Full-sky searches for nearby, faint objects are an obvious way to search for Mirror Stars. An object about $10^{-3}$ times the luminosity of the sun (not an unreasonable figure for a value of $\epsilon$ between $10^{-10}$ and $10^{-11}$) could be detected by Gaia within approximately 1000 light years. Other benchmarks can be much dimmer and could likely only be seen if they were significantly closer. The X-ray luminosity of our Mirror Stars is likely too dim to be detected by an X-ray full sky search. More promising is the possibility that, once candidates have been identified in optical surveys, an X-ray telescope such as Chandra could be pointed in the same direction with a long exposure. Taking Chandra's best, long exposure sensitivity as a benchmark (as achieved during its survey of the Hubble Deep Field North, approx. $4.9\times10^{-17}$~erg~m$^{-2}$~s$^{-1}$ in the 0.5-2 keV band, and $2.3\times10^{-16}$~erg~m$^{-2}$~s$^{-1}$ in the 2-8 keV band \cite{Brandt:2001vb}), we find that the X-ray signal could also be resolved a little over 100 light years away, depending on the benchmark.

\begin{figure}[t!]
\centering
\includegraphics[scale=0.6]{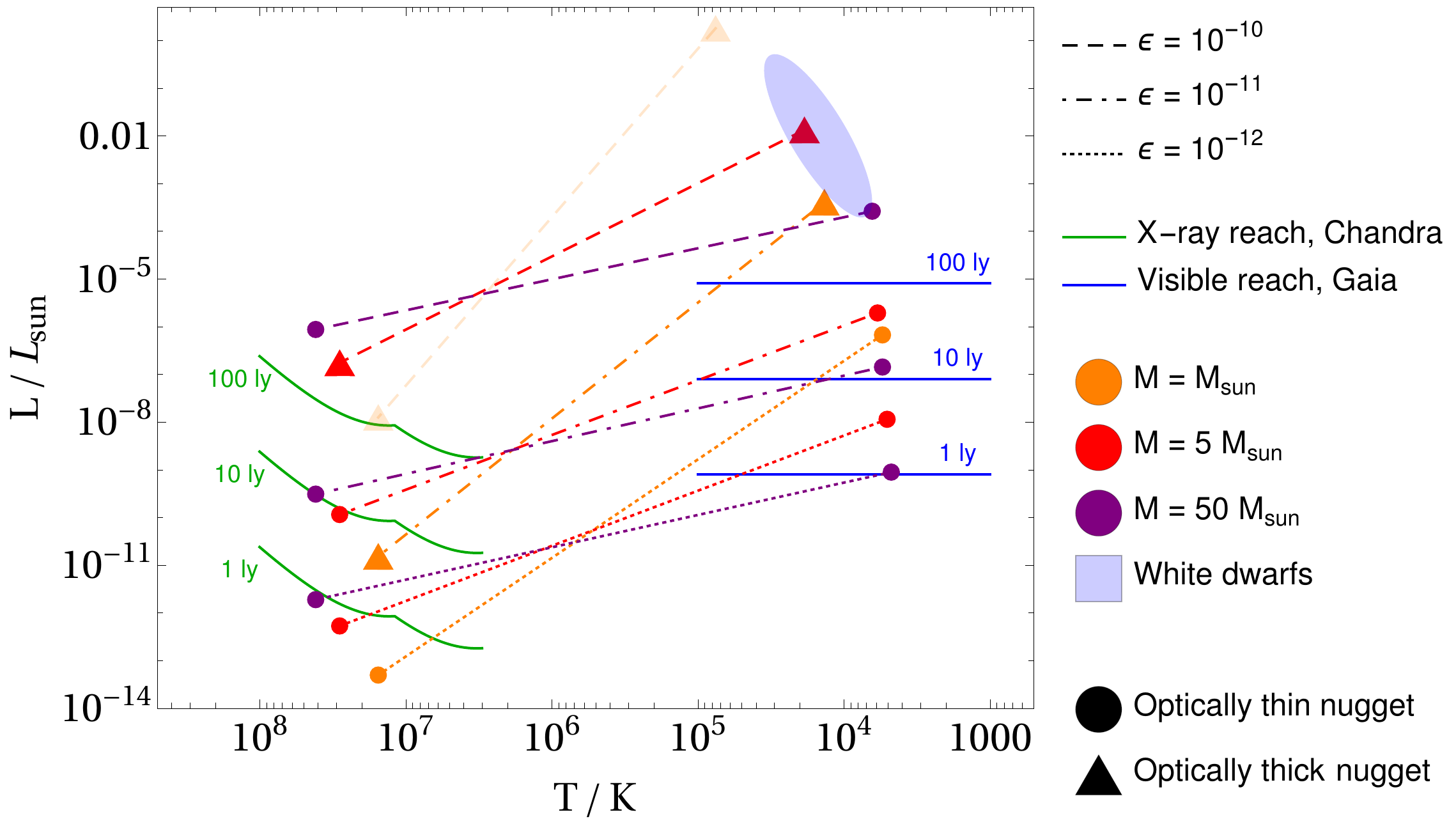}
\caption{A Hertzsprung-Russell diagram showing the dual signatures of our Mirror Star benchmarks. The plot shows luminosity against temperature, and each star is represented by two points connected by a line, each star having both a thermal emission in visible frequencies and an X-ray emission. The solid lines show the approximate distances up to which such objects could be observable via different observational techniques. The observability of the  visible frequency signal is compared to Gaia's reach for objects of the same absolute magnitude, and the observability of the X-ray signal is compared to the limiting sensitivity of the Chandra X-ray Observatory. The kink in the X-ray sensitivity lines results from us selecting whichever of the two X-ray energy bands has a better sensitivity for a black body signature at that temperature. One of the points has been made transparent; this is the benchmark for which the luminosity of the nugget exceeds that of the star, signaling that our assumptions of ignoring the effect of the SM nugget on the Mirror Star itself have broken down.}
\label{figure:HRplot}
\end{figure}

\begin{figure}[t!]
\centering
\includegraphics[scale=0.6]{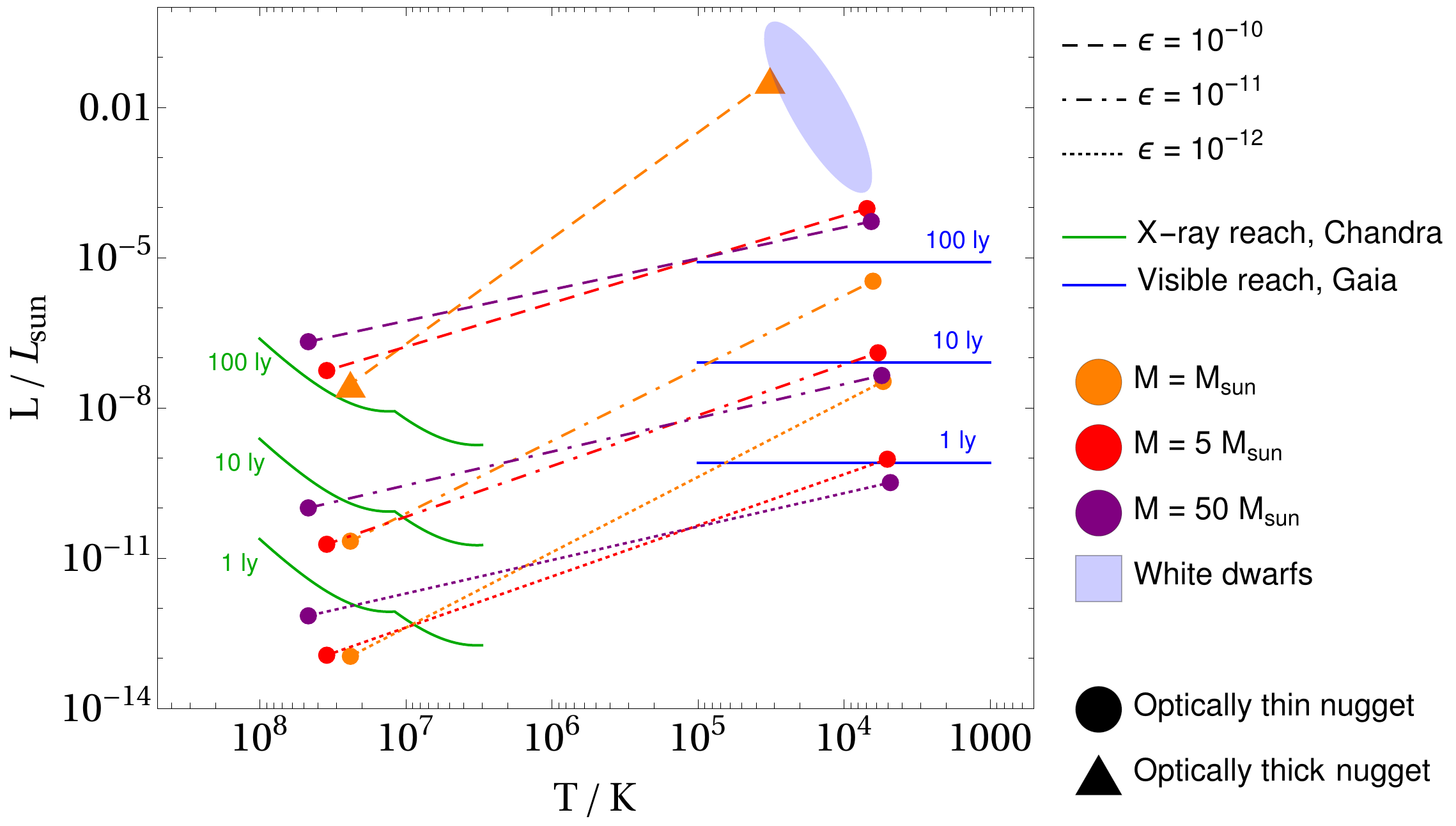}
\caption{A Hertzsprung-Russell diagram showing the dual signatures of our helium rich Mirror Star benchmarks. Labeling same as Figure~\ref{figure:HRplot}.}
\label{figure:HRplot_helium}
\end{figure}

In Figure \ref{figure:HRplot} we plot the signatures for a variety of benchmarks Mirror Stars on a Hertzsprung-Russell-type diagram. For each star we show both the visible and the X-ray signals, connected by a line, along with an indication of how far away we expect such luminosities to be observable. Benchmarks with optically thin nuggets are denoted by circles, and those with optically thick nugget by triangles. 

For a solar mass Mirror Star with $\epsilon = 10^{-10}$, the star captures so much material that the heating rate calculated according to Section~\ref{section:heating_and_cooling} is higher than the overall luminosity of the Mirror Star. Taken seriously, this would imply that the nugget is able to draw heat from the Mirror Star core so efficiently that it is able to `quench' the star, clearly violating our assumption that the accumulated SM matter does not significantly influence the Mirror Star.
Furthermore, the SM nugget is optically thick in these cases, and it is  possible that radiative heat transport is inefficient enough that the inner layers of the nugget thermalize with the Mirror Star core. In that case our calculation of the heating rate in Section~\ref{section:heating_and_cooling} is no longer reliable, since it assumes that heat flows only from hot mirror matter to much cooler SM matter. 
This is particularly interesting because it suggests that accumulation of mirror matter in our sun could lead to observable deviations from the standard model of solar physics, supplying new bounds on theories of Neutral Naturalness and other models with an asymmetric dissipative dark matter component.
An analysis of these more extreme scenarios is currently in preparation.

\subsection{Generality of Results}

Despite the lengthy calculations we have outlined in this paper, the basic principle of the signatures of Mirror Stars is rather simple. The Mirror Star captures hydrogen and helium from the interstellar medium, which falls into the hot, dense core and is heated up by $\epsilon^2$-suppressed photon portal interactions with the extremely hot Mirror Star core material. 
The nugget effectively emits two thermal signals.
The first is associated with its own equilibrium temperature, which fairly model-independently will be $\mathcal{O}(10^4 \mathrm{K})$. This is the temperature where the captured SM hydrogen and helium starts becoming slightly ionized and highly effective cooling mechanisms kick in, such that the equilibrium temperature only depends weakly on the heating rate.\footnote{It is no coincidence that this is also the surface temperature range of regular stars, since photons can leave the interior of the star once they reach a region where the degree of ionization drops sufficiently to allow escape from the star.}
The second thermal signal is set by the core temperature of the Mirror Star and arises from mirror-X-ray conversion into SM X-rays.

We can therefore anticipate the outcome of applying our signal calculation to non-SM-like Mirror Stars, i.e. those formed of mirror particles with masses different than their SM counterparts, as would arise in theories of Neutral Naturalness~\cite{Chacko:2016hvu,Craig:2016lyx,Chacko:2018vss, MTHastro}. 
The captured nugget will have an equilibrium temperature in the $\mathcal{O}(10^4 \mathrm{K})$ range and radiate accordingly. The frequency of the mirror-X-ray conversion signal will be determined by the Mirror Star core temperature.  This determines the key features of the shape of the visible Mirror Star spectrum, although the  luminosity  of the mirror-X-ray conversion signal depends on the opacity of the SM nugget and needs to be calculated as shown in Section~\ref{section:xrayconversionsignal}.

To estimate the luminosity of the SM photon signatures, we need to estimate the total size of the SM nugget, i.e. the number $N_{SM}$ of SM atoms that accumulate in the Mirror Star. This  depends on the lifetime of the Mirror Star $\tau_{star}$, which depends on the rate that the star burns its fuel, so some knowledge of the details of the mirror nuclear physics will be required. We saw in equation \eqref{equation:overall_rate} that the amount captured (in the non-geometric regime) also depends on the total number of mirror nuclei targets in the star, or equivalently its mass $M_{star}$, and scales with $\epsilon^2$. Thus the total signal strength scales with $N_{SM} \sim \epsilon^2 \tau_{star} M_{star}$.

The signal strength also depends on the rate at which the captured SM matter is heated, see \eqref{equation:real_heating}. For the Mirror Star benchmarks we have studied, the radius of the nugget is always very small compared to the radius of the star, so that the properties (temperature, density) of the mirror stellar matter do not change appreciably over the volume of the nugget. This means that a reliable estimate of the heating rate can be obtained from just the core temperature $T_{mirror}^{core}$ and core density $\rho_{mirror}^{core}$ of the Mirror Star, as well as the masses and charges of mirror nuclear species. 
We therefore see that  the total signal strength, set by the total heating rate at equilibrium, scales as
\begin{equation}
P \  \sim \  \epsilon^4 (T_{mirror}^{core})^{-1/2} \rho_{mirror}^{core} \tau_{star} M_{star}
\end{equation}
where we already incorporated the linear dependence on the nugget size $N_{SM}$ in the non-geometric regime.

The signal strength thus scales almost trivially with the Mirror Star mass, lifetime, core density and core temperature -- although there is the interesting possibility that if Mirror Star cores were significantly hotter than Standard Model stars, the X-ray signal (which scales with $T^4$) could end up dominating over the thermal nugget emission signal.

We will not comment on how these Mirror Star properties would be expected to depend on the Lagrangian of the mirror sector, as this would call for a detailed stellar physics analysis and is beyond the scope of this paper, but once these properties are obtained, applying our analysis to these exotic Mirror Stars is straightforward.

\section{Conclusions and Outlook}
\label{section:conclusions}

Mirror Stars are a generic prediction of complex dark sectors containing analogues of nuclear physics and electromagnetism, and arise in well-motivated models that address the hierarchy problem, like the asymmetrically reheated Mirror Twin Higgs framework ~\cite{Chacko:2016hvu,Craig:2016lyx,Chacko:2018vss, MTHastro}. 
In this paper, we study their astrophysical signatures for the first time.

We show that if the dark and SM photon have a small kinetic mixing $\epsilon < 10^{-9}$, Mirror Stars capture SM matter from the interstellar medium in their cores. 
This SM nugget is heated up by highly suppressed interactions with the mirror matter, giving rise to a thermal emission signal at $T \sim 10^4$ K at optical and IR frequencies. 
The amount of captured SM matter, and hence the total SM photon luminosity, depends simply on the size and age of the Mirror Star, as well as fundamental parameters of the hidden sector.
The captured SM matter also acts as a catalyst for mirror Thomson conversion of thermal mirror X-rays in the Mirror Star core. A fraction of these X-rays escapes the SM nugget and can be observed at frequencies $\sim T_{core}$, providing a direct window into the Mirror Star interior and the dark nuclear processes that reign within.

This double signature is highly distinctive. In optical surveys, Mirror Stars would look similar to white dwarfs, but likely with an absolute luminosity much too low to be consistent with a known astrophysical stellar object at the observed temperature.
This itself is a remarkable signature, but a dedicated follow-up X-ray observation can then detect the X-ray conversion signal, providing the smoking gun of a Mirror Star. 
For $\epsilon \sim 10^{-12} - 10^{-10}$, we find that optical surveys like Gaia and X-ray observatories like Chandra could discover Mirror Stars at distances up to 100-1000 light years away.

At higher photon mixings $\epsilon \gtrsim 10^{-10}$, Mirror Stars might look superficially like White Dwarfs, though detailed spectral analysis is still likely to uncover inconsistencies. In this case, the X-ray signal would still provide conclusive evidence of the Mirror Star's nature, providing additional motivation to study white dwarfs with X-ray observations, see also~\cite{Dessert:2019sgw}. 
Kinetic mixing values in this range could also be constrained by mirror capture in the Sun, depending on the density of the ``mirror interstellar medium''.

Our investigation used SM-like Mirror Stars as a benchmark. 
We showed that estimation of their signal as a function of dark photon kinetic mixing $\epsilon$ only requires knowledge of their rough mass, lifetime, and core temperature, in addition to the fundamental parameters of the hidden sector. 
This makes it straightforward to apply our techniques to Mirror Stars arising in more general or motivated hidden sector theories like Neutral Naturalness, which we will investigate in the future. 
An understanding of mirror stellar astrophysics will also allow us to study other signatures, like mirror supernovae, or mirror stellar relics that could show up in gravitational wave observations. 

In conclusion, we have shown that Mirror Stars generate highly distinctive and discoverable signals that provide robust windows into the underlying hidden sector physics. 
The discovery potential is impressive, and deserves dedicated observations to open up a new fundamental frontier into our universe's dark matter sector.

\textbf{Acknowledgements:} 
We  thank 
Christopher Matzner,
 Zackaria Chacko, Christopher Dessert, Michael Geller, Bob Holdom,
 Yoni Kahn, Benjamin Safdi, 
 Kai Schmidt-Hoberg,
 and Yuhsin Tsai
   for helpful conversations. DC would like to especially thank Zackaria Chacko, Michael Geller and Yuhsin Tsai for early discussions on the possibility of Mirror Stars in MTH models.
 The research of DC and JS was supported by a Discovery Grant from the Natural Sciences and Engineering Research Council of Canada.

\bibliographystyle{JHEP}
\bibliography{References}

\end{document}